# Microstructure, mechanical properties, corrosion resistance and cytocompatibility of WE43 Mg alloy scaffolds fabricated by laser powder bed fusion for biomedical applications

Muzi Li[1], Felix Benn[2,3], Thomas Derra[3], Nadja Kroeger[4], Max Zinser[4], Ralf Smeets[5], Jon M. Molina-Aldareguia[1], Alexander Kopp[3], Javier LLorca[1,6]

[1]*IMDEA Materials Institute, C/Eric Kandel 2, Getafe, Madrid, 28906, Spain*
[2]*Queen's University Belfast, University Road, Belfast, BT7 1NN, Northern Ireland, UK*
[3]*Meotec GmbH & Co. KG, Philipsstr. 8, 52068 Aachen, Germany*
[4]*University Clinic Cologne, Kerpenerstr. 62, 50931 Köln, Germany*
[5]*University Medical Center Hamburg-Eppendorf, Martinistr, 52, 20246 Hamburg, Germany*
[6]*Department of Materials Science, Polytechnic University of Madrid, 28040 Madrid, Spain*

**Abstract**

Open-porous scaffolds of WE43 Mg alloy with a body-center cubic cell pattern were manufactured by laser powder bed fusion with different strut diameters. The geometry of the unit cells was adequately reproduced during additive manufacturing and the porosity within the struts was minimized. The microstructure of the scaffolds was modified by means of thermal solution and ageing heat treatments and was analysed in detail by means of X-ray microtomography, optical, scanning and transmission electron microscopy. Moreover, the corrosion rates and the mechanical properties of the scaffolds were measured as a function of the strut diameter and metallurgical condition. The microstructure of the as-printed scaffolds contained a mixture of Y-rich oxide particles and Rare Earth-rich intermetallic precipitates. The latter could be modified by heat treatments. The lowest corrosion rates of 2-3 mm/year were found in the as-printed and solution treated scaffolds and they could be reduced to ~0.1 mm/year by surface treatments using plasma electrolytic oxidation. The mechanical properties of the scaffolds improved with the strut diameter: the yield strength increased from 8 to 40 MPa and the elastic modulus improved from 0.2 to 0.8 GPa when the strut diameter increased from 275 µm to 800 µm. Nevertheless, the strength of the scaffolds without plasma electrolytic oxidation treatment decreased rapidly when immersed in simulated body fluid. In vitro biocompatibility tests showed surface treatments by plasma electrolytic oxidation were necessary to ensure cell proliferation in scaffolds with high surface-to-volume ratio.

# 1 Introduction

Critical-sized bone defects are defined as those that will not heal spontaneously within a patient's lifetime [1-2]. They are often sequelae of high-energy impact or trauma, tumor resection, infection, osteo-degenerative diseases or revision surgery [3-4]. A further reason for loss of bone may be congenital bone anomalies, which are usually due to the absence or maldevelopment of bone [4]. Because of the load bearing requirements, traditional implants for bone healing are made of Ti alloys, stainless steel or cobalt alloys. They are non-biodegradable and, therefore, a second surgery is often needed to remove the implants once their purpose is fulfilled because they might lead to endothelial dysfunction, permanent physical irritation/or and chronic inflammatory local reactions [5]. These limitations may be overcome with the use of bioabsorbable metals that are able to degrade or corrode gradually *in vivo*. The corrosion products can be eliminated and/or metabolised or assimilated by cells and/or tissue and the implant dissolves completely after tissue healing with no implant residues [6].

Among the biodegradable metals, Mg presents the best potential for bone implants because of its biocompatibility and osteopromotive properties which can stimulate new bone formation [7-9], while the excess of Mg ions generated during biodegradation can be safely eliminated via the circulatory system and excreted by urine. In addition, the elastic modulus of Mg (~45 GPa) is similar to that of human bones (2-30 GPa) [10], which can reduce stress shielding during load transfer at the bone-implant interface [11-12]. As a result, Mg alloys are currently used in different orthopaedic implants in the form of screws, plates, nail supports and wires for orthopaedic surgeries [13-14].

One further development in the application of Mg alloys for bone healing is the fabrication of porous scaffolds that can be loaded with the appropriate osteogenic cells and growth factors in order to accelerate the growth of the new tissue. Manufacturing of porous Mg scaffolds by means of powder metallurgy or laser machining was demonstrated in the past [15-16] but it became feasible very recently with new developments in additive manufacturing. Open-porous scaffolds of Mg alloys can be fabricated by laser powder bed fusion (LPBF) with controlled topology and pore structure (size, shape and connectivity) together with customised geometry that matches the anatomy of patients [17]. Jauer *et al.* were the first to manufacture Mg scaffolds with minimum porosity within the struts [18]. Afterwards, Li *et al.* analysed the *in vitro* properties of WE43 Mg alloy scaffolds formed by diamond cells [19]. They found that the cytotoxicity levels were below 25%, the compressive strength was within those reported for trabecular bone after 4 weeks of degradation in simulated body fluid (SBF) while the volume loss after 4 weeks was around 20%. A follow-up study of the same scaffolds showed that biodegradation decreased the fatigue strength of the porous material from 30% to 20% of its yield strength and that cyclic loading significantly increased its biodegradation rate. [20]. Kopp *et al.* [21] also studied the mechanical properties and biodegradation of WE43 Mg alloy scaffolds formed by square cells with different pore size and with or without the modification of the surface by plasma electrolytic oxidation (PEO). They found that the surface modification by PEO improved the corrosion resistance in Dulbecco's Modified Eagle Medium (DMEM), especially at the initial stages, and the mechanical properties after exposure to this medium. The PEO surface modification was also beneficial for increasing cell affinity and growth, which in turn resulted in better bone-scaffold integration [22].

Fast biodegradation is known to be on the main challenges with Mg implants because of the rapid release of hydrogen and the premature failure of the implants. This limitation may be even more critical for porous scaffolds manufactured by LPBF due to the enlarged surface area, complex geometry and metastable microstructure generated during rapid solidification. In fact, the microstructural features of WE43 Mg alloys manufactured by casting and LPBF show very different texture and precipitate distribution that can dramatically influence the mechanical properties and corrosion resistance [23].

Therefore, the objective of this investigation was to provide a deeper understanding of the effect of the scaffold geometry, microstructure and surface treatments on the mechanical properties, corrosion resistance and cytocompatibility of WE43 MG alloys scaffolds manufactured by LPBF. To this end, cubic scaffolds with three different strut and pore sizes were manufactured and the microstructure was modified by means of thermal treatments [24-28]. The mechanical response in compression along the building direction and perpendicular to the building direction was determined by means of *in situ* tests within either optical or scanning electron microscopes to ascertain the deformation and failure mechanisms before and after immersion in SBF, while direct and indirect cytocompatibility tests were carried out to assess the influence of the surface treatments by PEO on cell proliferation.

## 2. Material and experimental techniques

*2.1 Sample preparation by LPBF manufacturing of porous scaffolds and surface treatment*

Mg alloy WE43 MEO (Meotec GmbH, Aachen, Germany) with a nominal composition of 1.4 – 4.2 % Y, 2.5 - 3.5 % Nd, <1 % (Al, Fe, Cu, Ni, Mn, Zn, Zr) and balance Mg (in wt%) was casted and powdered for powder metallurgical processing. The powder particle distribution was $D_{10}$ = 29.1 µm, $D_{50}$ = 45.8 µm and $D_{90}$ = 64.4 µm [21]. The particles are naturally covered with a thin passivation layer of $Y_2O_3$ which also limits the flammability of the powder during handling. Alloys with similar composition have been successfully applied in orthopaedics applications [29] and cardiovascular stents [30] and are well-established as bio-degradable and bio-compatible materials [6].

Cubic open-porous scaffolds of 10×10×10 mm³ based on a unit cell with a body-centered cubic structure (cylindrical struts join cube corners with adjacent corners as well as with adjacent cube centers) were designed with three different nominal strut diameters of 250 µm, 500 µm and 750 µm. They will be denominated hereafter small strut (SS), medium strut (MS) and large strut (LS) scaffolds, respectively (Table 1). The nominal pore cell size increased from 720 µm to 860 µm as the strut diameter increased (Table 1). They were manufactured using the LPBF method in an Ar atmosphere using an Aconity mini 3D printer with a maximum laser power < 400 W and a beam diameter of 90 µm using a 0º/90º scan strategy for printing successive layers. The approximate thickness of each printed layer was 30 µm. In addition, a bulk cylindrical sample of 8 mm in diameter and 30 mm in length was manufactured using the same processing parameters in order to prepare electro-polished samples for transmission electron microscopy (TEM). The as-printed scaffolds were etched in 50% concentrated phosphoric acid solution for 5 min to remove the excess material at the surface due to sintering process. The specimen surface of several scaffolds was modified by PEO using a phosphate-based electrolyte (Kermasorb®) to achieve a continuous surface coating with an average thickness of 9 ± 6 µm. More details about the LPBF process and the PEO surface modification can be found in [21].

Table 1. Geometrical parameters of the scaffolds with different strut diameter.

|  | Nominal strut diameter (µm) | Nominal pore size (µm) | Actual strut diameter (µm) | Actual pore size (µm) | Surface area (mm$^2$) | Porosity (%) | Density (g/cm$^3$) |
|---|---|---|---|---|---|---|---|
| **SS** | 250 µm | 720 | 275 ± 28 | 718 ± 24 | 3232 | 76.7±2.1 | 0.43±0.04 |
| **MS** | 500 µm | 830 | 518 ± 27 | 843 ± 36 | 2210 | 63.1±1.6 | 0.68±0.03 |
| **LS** | 750 µm | 860 | 793 ± 39 | 850 ± 35 | 2207 | 58.2±2.8 | 0.77±0.05 |

*2.2 Heat treatments*

Two different heat treatments were selected to investigate their influence on the microstructure, mechanical properties and corrosion resistance in comparison with the as-printed scaffolds. The first one, denominated T4, involved a solution treatment at 525°C for 8 h, followed by water quenching to create a supersaturated solid solution of the alloying elements into the Mg matrix. The second treatment, denominated T6, included an ageing step at 250°C for 16 h after the T4 treatment to additionally create a homogeneous dispersion of nm-sized precipitates. This treatment has been reported as to provide peak hardness for cast WE43 alloys [24].

*2.3 Microstructure characterisation and chemical composition*

Metallographic samples were prepared from the scaffolds by manually grinding with SiC abrasive paper 4000 grit, followed by mechanical polishing with diamond paste of 3 µm and afterwards of 0.25 µm using an alcohol-based lubricant to avoid oxidation. Two different etchants were applied on the polished surfaces: 5% nital was used to reveal laser scan tracks under the optical microscope (Olympus BX51) while a mixture of 75 ml ethylene glycol, 24 ml distilled water and 1% nitric acid was used to observe intermetallic phases under the scanning electron microscope (SEM). Electron backscattered diffraction (EBSD) analysis was conducted at 30 kV and 0.69 nA probe current with a step size of 0.3 µm. The grain size and inverse pole figures were obtained by post processing the EBSD map with the software Oxford® HKL Channel-5.

TEM samples were prepared from two different routes. Thin plates of 200 µm thickness were cut parallel and perpendicular to the building direction from the cylindrical bulk specimen with a slow speed cutting blade. Discs of 3 mm in diameter were mechanically punched from these plates and manually polished down to ~80 µm. The thinned discs were twin jet polished with an etchant of 10.6 g lithium chloride, 22.32 g magnesium perchlorate, 200 ml 2-butoxy-ethanol and 1000 ml methanol at -40 °C and 95 V. For the microstructural analysis of aged scaffolds, TEM samples were prepared by focused ion beam milling in a FEI Helios NanoLab 600i dual-beam electron microscope down to electron transparency (<100 nm). TEM observation was carried out in a FEI Talos equipped with a field emission gun operating at 200kV. Both bright field (BF) images in TEM mode and high-angle annular dark field (HADDF) images in scanning-transmission electron microscopy (STEM) mode were acquired. Chemical composition mapping was performed under STEM mode with a SuperX EDX detector. The acquisition time was set between 30-60 min and data was analysed using Bruker QUANTAX software.

*2.4 Corrosion tests*

Corrosion tests were carried out on scaffolds that were not protected by PEO surface modification to study the immediate influence of the alloy microstructure in different conditions on the degradation rates. The degradation medium was DMEM with high glucose (4500 mg/L glucose, phenol red, NaHCO₃) and the degradation rate was determined by measuring the hydrogen release using an experimental set-up described in [21] or from the X-ray computed tomography (XCT) tomograms. The chemical reaction of one mole of Mg corresponds to the release of one mole of hydrogen [25] and thus, the average corrosion rate *CR* (expressed in mm/year) can be determined assuming that the mass loss $\Delta M$ is uniformly distributed on the exposed surface area of scaffold *A* according to Eq. (1):

$$CR = K\frac{\Delta M}{At\rho} \tag{1}$$

where $K= 8.76 \times 10^4$, $\Delta M$ is expressed in $g$, A in $cm^2$, $t$ is the immersion time in $h$, and $\rho$ the density in $g/cm^3$.

*2.5 X-ray tomography*

The WE43 scaffolds with different strut diameters were examined by XCT before and after the degradation tests in a Phoenix Nanotom® with a tungsten target and a filter of 0.2 mm Cu with slit collimator. The sample-focus distance was set to 25 mm and the detector-focus distance to 200 mm. Optimised operating parameters include 60 µA tube current, 120 kV tube voltage, 6.25 µm per pixel resolution and 1900 projections during rotation of the sample. The scan time to obtain the tomogram for each scaffold was about 2.5 hours. The strut diameter and pore size were determined from 3D tomograms using ImageJ®, the latter using the intersection method (Fig. 2b). The total surface area was calculated from 3D tomograms using Avizo® software.

*2.6 Mechanical characterisation*

Compression tests of the cubic scaffolds (before and after degradation) were carried out using a Kammrath Weiss GmbH® micromechanical testing machine at a crosshead speed of 0.6 mm/min up to a maximum engineering strain of 60%. Videos were recorded upon deformation using an optical microscope. One compression test was also performed *in situ* within the SEM, and secondary electron micrographs were taken at intervals of 1% strain. The mechanical properties of the porous scaffold were determined from the load and displacement data following the procedures indicated in ISO 13314:2011 for porous structures with porosity > 50%. The nominal stress *S* was determined from the load divided by the nominal cross-section of the scaffolds while the engineering strain ε was calculated as the cross-head displacement of the mechanical testing machine divided by the initial height of the sample. Tests were carried out in PEO-modified scaffolds with different strut diameters in orientations parallel and normal to the building direction to assess the influence of these factors on the deformation and fracture mechanisms. In addition, the influence of the heat treatment and of the immersion in SBF on the mechanical properties was measured along the building direction in scaffolds without PEO surface treatment.

*2.7 In vitro cytocompatibility*

L-929 mouse fibroblasts (LGC Standards, Wesel, Germany) were cultured in Minimum Essential Medium (MEM) supplemented with 10% foetal bovine serum, penicillin/streptomycin (100U/ml each) at 37°C, 5 % $CO_2$ and 95% humidity. RM-A, a polyurethane film containing 0.1 % Zinc diethyldithio-141 carbamate (ZDEC), was used as the positive control reference material for the indirect and direct assays. Toluene resisting plastic sheets were utilized as negative non-toxic controls for the direct assays.

2.7.1 Indirect test

The extracts were obtained by incubating the samples in MEM with SBF at a ratio of 1 ml / 3 $cm^2$ for 72 h under the previously mentioned cell culture conditions. Cell culture medium without samples was used as the negative control extract. After incubation, the extracts were centrifuged at 14,000 rpm for 10 min. The supernatants were utilized for the indirect biocompatibility assays described below. L929 cells were seeded with $1 \times 10^4$ cells/well in 100 μL cell culture medium and incubated for 24 h in 96 well plates. Then the medium was discarded and replaced with 100 μL of extract. Cells were incubated for another 24 h and then immediately subjected to the Bromodeoxyuridine/5-Bromo-2-Deoxyuridine (BrdU) and Sodium 3,3′-[1(Phenylamino)Carbonyl]-3,4-Tetrazolium]-3is(4-Methoxy-6-Nitro) Benzene Sulfonic acid Hydrate (XTT) assays, whereas the corresponding supernatants were subjected to the Lactate Dehydrogenase (LDH) assay. Cell culture medium and extracts without cells respectively were served as the blank controls and were thus subtracted from the measured absorbance values.

For the BrdU assay, a standard colorimetric test kit (Roche Diagnostics, Mannheim, Germany) was utilized according to the manufacturer's instructions. In brief, cells were labeled with BrdU under cell culture conditions and subsequently fixed at RT with FixDenat reagent. The cells were then incubated with anti-BrdU-peroxidase (POD) antibody and washed. Immune complexes were then detected via subsequent substrate reaction with added tetramethyl-benzidine (TMB) and following colorimetric measurement via multiwell scanning spectrophotometer with filters for 450 and 690 nm reference wavelength, respectively.

For the XTT assay, a standard test kit (Roche Diagnostics, Mannheim, Germany) was utilized according to the manufacturer's instructions. In brief, the electron-coupling reagent was mixed with XTT labeling reagent and 50 μL of this mixture was added to the cells. After incubation, the amount of substrate conversion was quantified by measuring the absorbance of 100 μL aliquots in a new 96 well plate via multiwell scanning spectrophotometer with filters for 450 and 650 nm reference wavelength, respectively.

For the LDH assay, a standard test kit (BioVision, Milpitas, USA) was utilized according to the manufacturer's instructions. In brief, cell supernatant was incubated with LDH reaction mix for 30 min at RT. Absorbances were measured via multiwell scanning spectrophotometer with filters for 450 and 650 nm reference wavelength, respectively.

2.7.2 Direct test

12 well plates were seeded with $2.4 \times 10^5$ cells in either 1 ml or 1.88 ml cell culture medium for surface-area to medium ratios of 5.65 or 3 $cm^2$/ml, respectively. Cells were seeded directly on the surfaces of the test specimens and incubated under previously mentioned cell culture conditions for 24 h before performing direct assay procedures.

For live-dead cell staining on the surfaces of the specimens, 60 ul Propidium iodide (PI) stock solution, consisting of Phosphate-Buffered Saline (PBS) supplied with 50 ug/ml PI were added for each ml of cell culture medium to each well. Additionally, 500 ul Fresh Fluorescein Diacetate (FDA) working solution, consisting of PBS supplied with 20 ug/ml FDA, which itself was in turn previously suspended in acetone stock solution at a concentration of 5mg/ml, were added for each ml of cell culture medium to each well. The wells were incubated for 3 min at RT. After sterilization by immersion in isopropanol for 5 min and subsequent drying on a sterile blanket underneath a laminar flow, the specimens were then rinsed in preheated PBS and immediately forwarded to upright red and green light fluorescence microscopy (Nikon Eclipse Ti-S/L100).

## 3. Results

*3.1 Geometry and chemical composition*

Optical micrographs of the WE43 scaffolds with different strut diameter (SS, MS and LS) and surface-modified by PEO are shown in Fig. 1a from left to right. The SS, MS and LS scaffolds contain 7×7×7, 5×5×5 and 4×4×4 unit cells, respectively. The average values and standard deviation of the strut diameter are shown in Table 1 and they were very close to the design values, taking into account the contribution of the PEO surface modification (thickness ≈ 10 µm). Thus, LPBF of Mg alloys scaffolds was very accurate, even for the thinnest struts. The structure of one body-center cubic unit cell within the scaffold is highlighted in Fig. 1b and the 3D reconstruction of LS scaffold by XCT is depicted in Fig. 1c. The surface of the periphery is relatively smooth with few un-melted or partially melted powder residues attached, while powder residues can be clearly observed inside of the pores from the 2D projections of the scaffold in Fig. 1d. Interestingly, less residues were found from the top view, where the pores going through the lattice are clearly visible. It is presumed that these could be removed with extra etching time, but this step would also result in more material loss, potentially weakening the scaffolds.

Videos S1 to S3 in the supplementary material show the XCT tomograms of the SS, MS and LS scaffolds at higher magnification. Two representative cross-sections perpendicular to the building direction are shown in Figs. 2a and 2b for the SS and LS scaffolds, respectively. The sections correspond to a plane equidistant from the center and surface of the cubic unit cells and the line to determine the pore size is depicted in Fig. 2b. They show that the geometry of the template was well reproduced by LPBF processing and that porosity within the struts was minimum. Powder residues were more evident on the SS scaffold, indicating that the strut dimensions were close to the limit achievable by LPBF. Two different grey levels were found throughout the scaffolds that stand for different materials with different X-ray absorption capacity. The oxide passivation layer on the powder particles is broken during the LPBF process and the oxide crusts - which are less likely to melt due to the high melting point - are mixed with the molten metal. These oxides contain Rare Earth (RE) elements such as Y and Nd, whose density is higher than that of Mg, leading to higher attenuation of the X-ray.

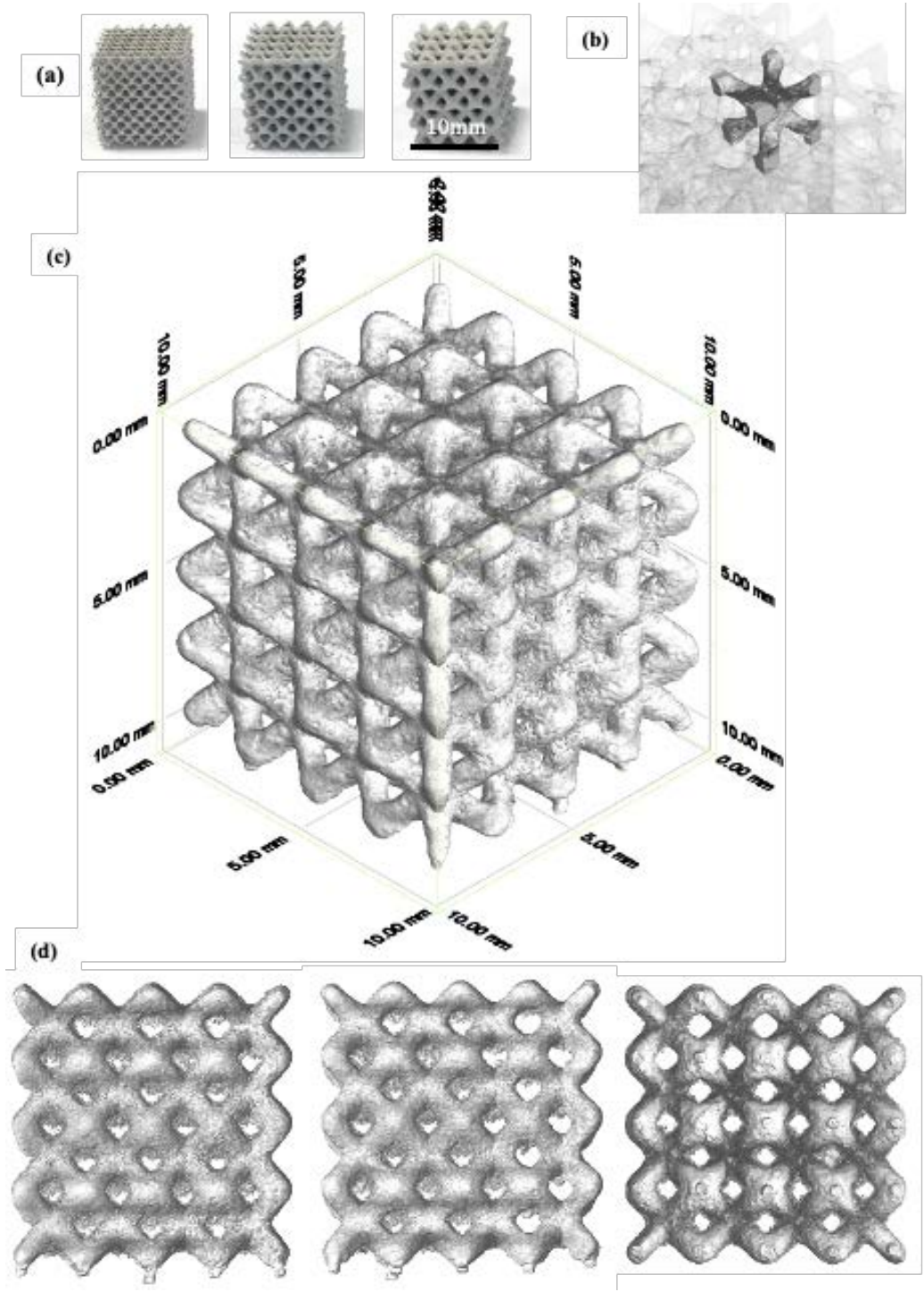

**Fig. 1.** (a) Images of PEO-modified scaffolds with different nominal strut diameter. Left to right: SS (250 µm), MS (500 µm) and LS (750 µm). (b) Highlight of one body-centre cubic unit cell in the scaffold (c) 3D reconstruction by XCT of the LS scaffolds with a step size of 6.25 µm. Building direction is vertical. (d) 2D front, left and upper views of (c) from left to right showing residual Mg powder agglomerates within the pores.

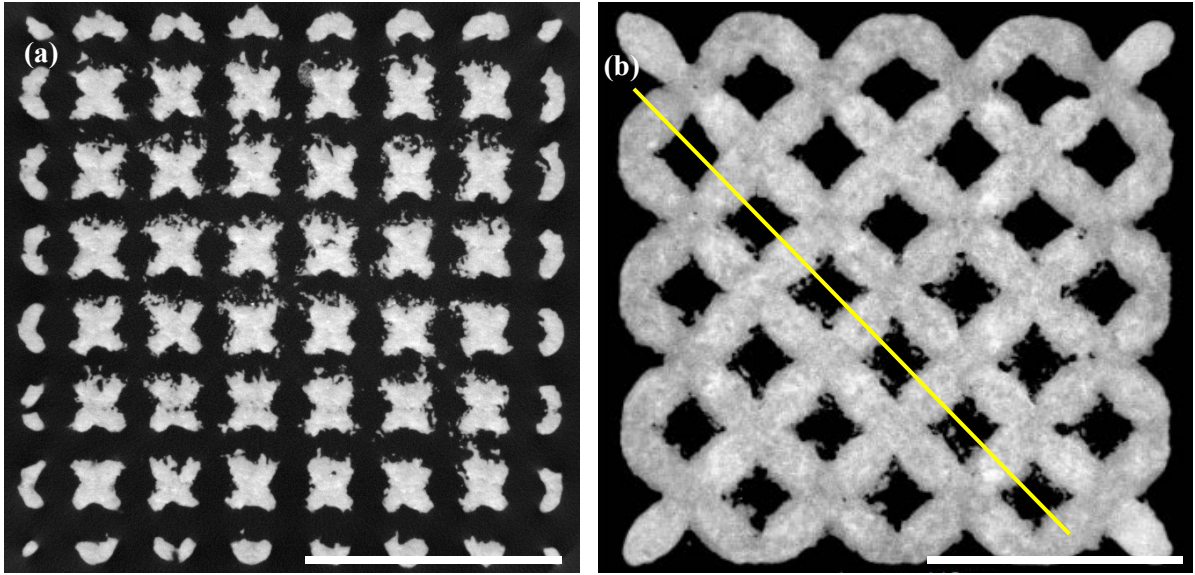

**Fig. 2.** High-resolution XCT cross-section of the scaffolds perpendicular to the building direction. (a) SS scaffold. (b) LS scaffold. The sections correspond to a plane equidistant from the center and surface of the cubic unit cells. The scale bar stands for 5 mm. The yellow line depicts the intersection method to measure the pore size.

*3.2 Microstructure*

3.2.1 As-printed alloy

The microstructures of the scaffolds with different strut diameter were analyzed and they were found to be equivalent. It should be noticed that the same processing parameters were used in all cases and the differences in strut diameter did not lead to noticeable microsutructural changes. The laser scan tracks delineated by the melt pool boundaries were clearly observed in the optical microscope after etching in nital (Fig. 3a). The melt pools were about 20 µm in depth and 70 µm in width, slightly smaller than the processing parameters (30 µm layer thickness and 90 µm focus diameter).

The microstructure of the ≈10 µm thick PEO modified surface layer is shown in the secondary electron (SE) image in Fig. 3b, where the dispersed oxide particles within the Mg alloy are also observed. The EBSD map of a cross-section of a strut parallel to the building direction is plotted in Fig. 3c, showing a bi-modal grain size distribution that contained a mixture of large grains (12.0 ± 6.1 µm) and small equiaxed grains with an average size of 2.1 ± 1.0 µm. The large

grains - which occupied most of the microstructure- were mainly oriented along the *c* axis, as indicated by the (0001) pole figure, which attains the maximum intensity near the rim. Thus, the *c* axes of the large Mg grains were oriented along the building direction while the small grains presented a random texture.

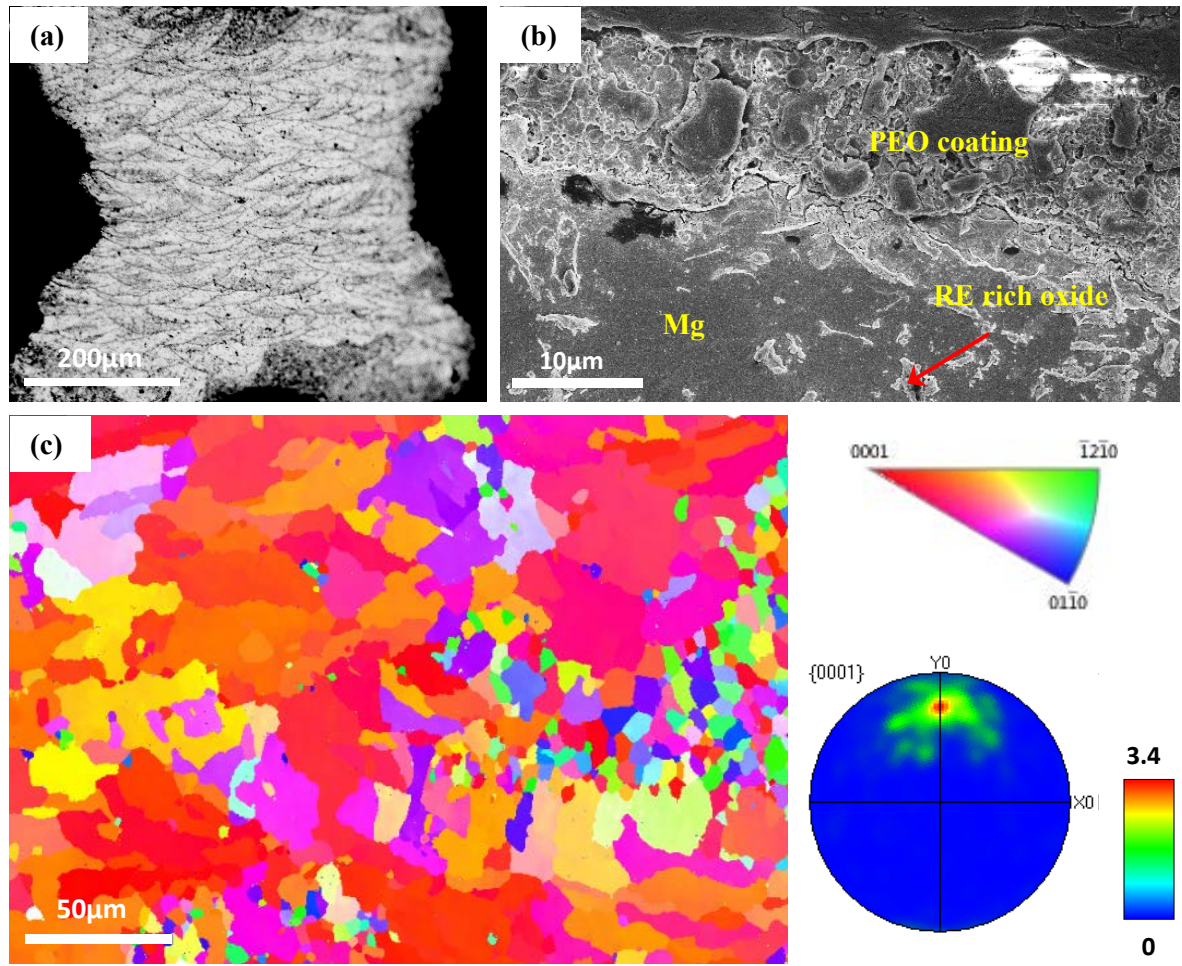

**Fig. 3.** (a) Optical micrograph showing the laser scan tracks in one strut of SS scaffold. (b) SE image showing the microstructure of the PEO surface modification. (c) EBSD map of the as-printed MS scaffold. The pole figure along the basal plane indicates a strong basal texture along building direction. The numbers in the legend stand for multiples of random distribution.

The general features of the microstructure of the WE43 Mg alloy processed by LPBF are shown in the backscattered electron (BSE) image in Fig. 4, where the laser scan tracks are visible. The main features in the Mg matrix are oxide particles from the powder crust and Mg-RE intermetallic precipitates. Both of them contain heavy RE and appear bright in the BSE images.

The oxide particles appear as large, irregular flakes dispersed in the grains of the dark Mg matrix with dimensions in the range 1 to 10 μm.

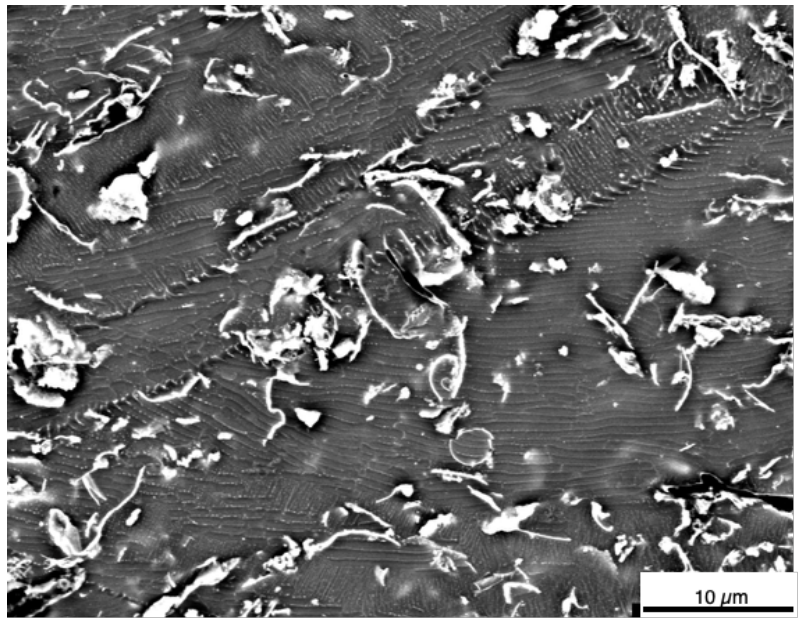

**Fig. 4.** BSE image showing the microstructure of the WE43 Mg alloy processed by LPBF. Yttrium oxides appear as large, irregular flakes with dimensions in the range 1 to 10 μm. Scan tracks are delineated by Mg-RE intermetallic precipitates, which also form horizontal lines (perpendicular to the building direction) as well as line perpendicular to the scan tracks. Building direction is vertical.

The intermetallic precipitates are much smaller and three different types (according to the size, shape and spatial distribution) can be distinguished depending on their location with respect to the melt pool boundaries (Fig. 5). One high-resolution BSE micrograph of a horizontal melt pool boundary is shown in Fig. 5a. One group of precipitates (marked with yellow arrows in Fig. 5a) have relatively large sizes and irregular shape and decorate the melt pool boundary. Other group of precipitates was found above the melt pool boundary forming vertical bands (parallel to the building direction) either in the form of wavy fibrous lines (Fig. 5b) or aligned spherical precipitates (Fig. 5c). Occasionally, both wavy fibrous and aligned round precipitates can be spotted in the same region. The distance between these vertical lines is around 0.36 ± 0.13 μm. Finally, the majority of the intermetallic precipitates were found in horizontal bands normal to the building direction below the melt pool boundaries and the band width was 0.52

± 0.10 μm (Fig. 5a). The precipitates in these horizontal bands also took two different forms: either long consecutive lines that sometimes contain nano porosity (Fig. 5d) or a series of round particles (<80 nm in diameter) forming bright lines in the BSE (Fig. 5e). It should be noted that the vertical and horizontal lines are more or less normal to each other in the same melt pool but they are not always strictly aligned or perpendicular to the building direction due to the fluctuations in the orientation of the thermal gradients during the LPBF process.

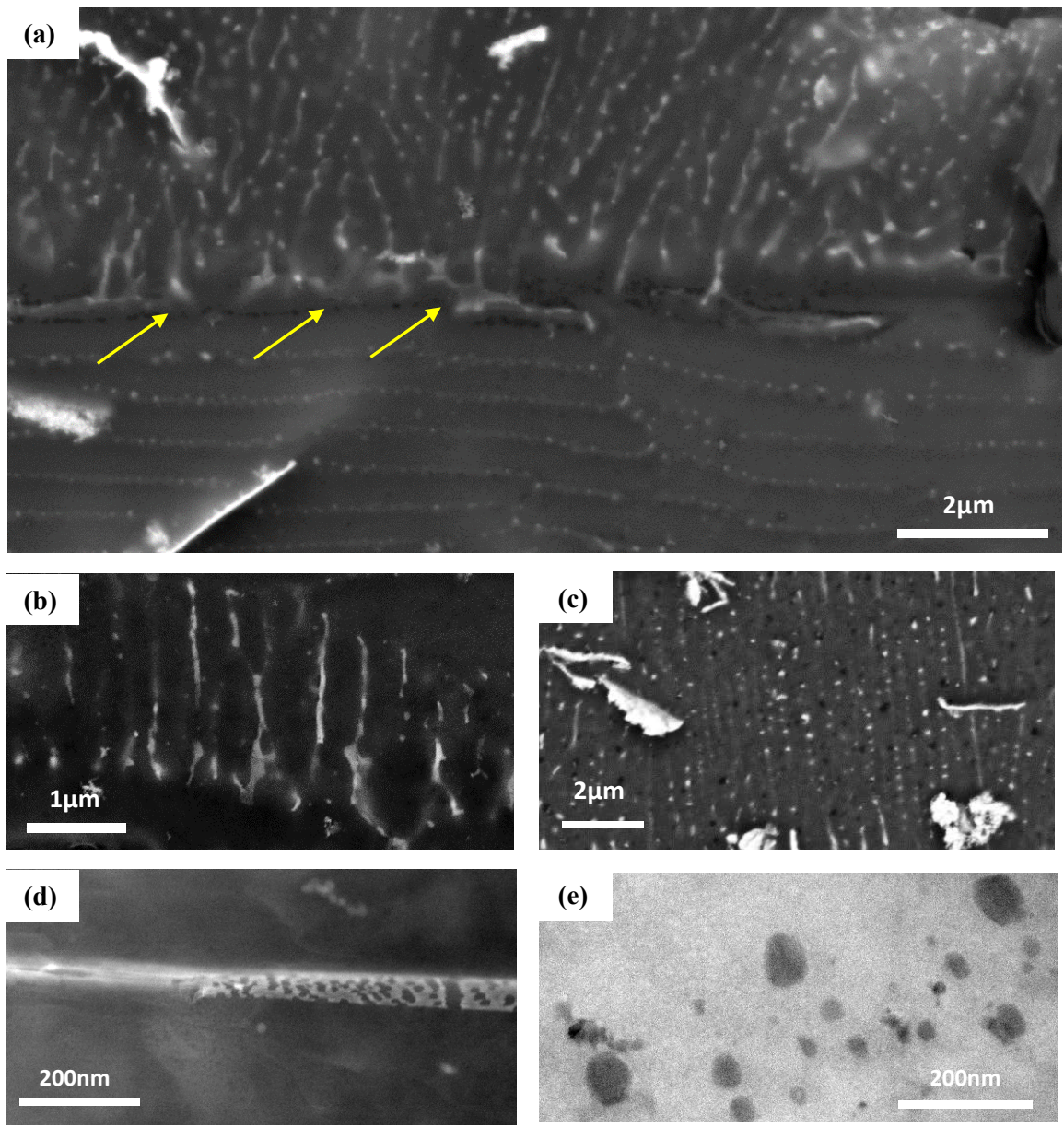

**Fig. 5.** (a) BSE image showing the three different precipitate structures in the as-printed WE43 Mg alloy scaffolds around the horizontal melt pool boundary. The yellow arrows mark the precipitates that decorate the melt pool boundary. (b) BSE image showing wavy fibrous lines of precipitates formed above the melt pool boundary. (c) BSE image showing aligned spherical precipitates formed above the melt pool boundary. (d) HAADF image showing a horizontal precipitate structure below the melt pool boundary (e) BF image showing horizontal lines formed by aligned round precipitates below the melt pool boundary. Building direction is vertical.

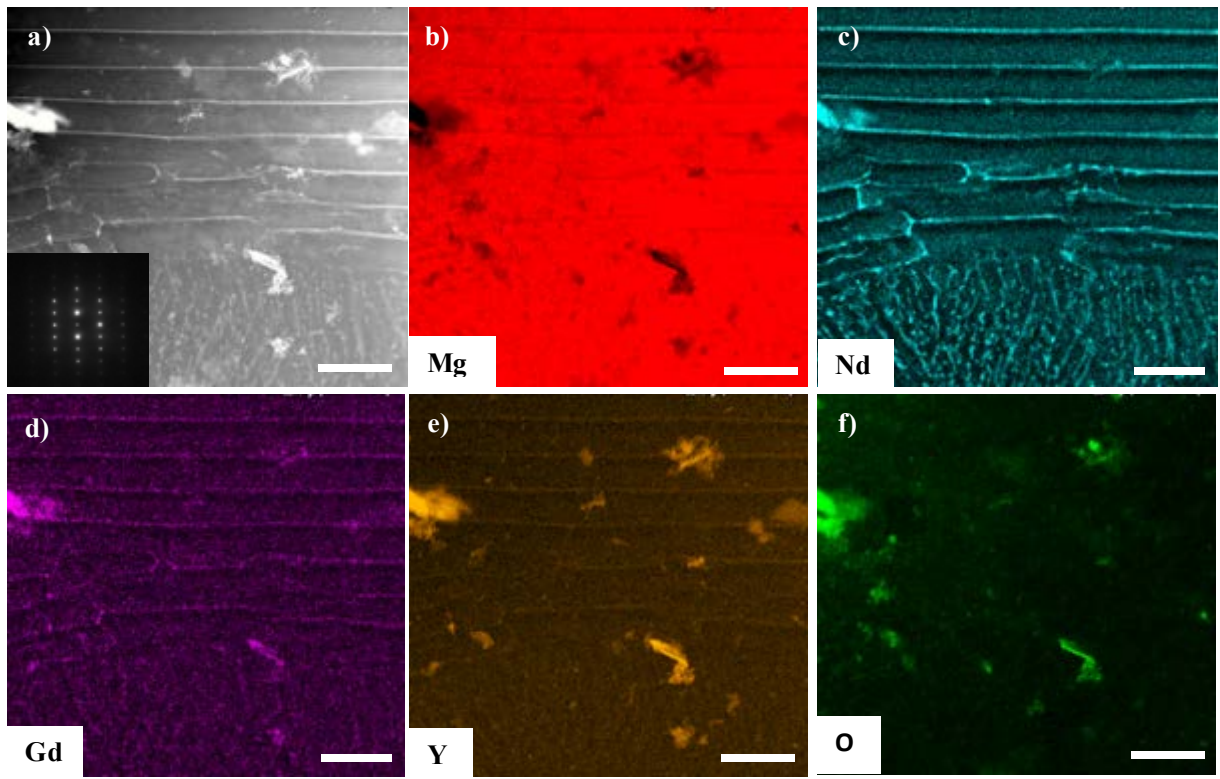

**Fig. 6.** (a) HAADF STEM image of the microstructure of a grain at the melt pool boundary. The corresponding SAED pattern of (10-10) zone axis reveals that the horizontal lines are parallel to the basal plane of Mg. (b-f) Content of the different chemical elements in (a) obtained by EDS. Building direction is vertical. The length of the white markers in each figure is 1 μm. Building direction is vertical.

The chemical elemental maps, obtained by EDS, in a grain at the melt pool boundary are plotted in Fig. 6. The grain is in the (10-10) zone axis, as shown by the selected-area electron diffraction (SAED) pattern in the inset of Fig. 6a, and the building direction is vertical. The μm-sized oxide particles with irregular shape are mainly formed by O and Y and also contain Gd, and Nd while they are depleted of Mg. Both vertical and horizontal lines formed by precipitates are clearly seen in the same melt pool. The horizontal lines are perpendicular to the building direction and are formed on the basal plane of the Mg matrix. These lines are mainly enriched in Nd and Gd (Fig. 6c and d) while Y (Fig. 6e) is less clear because the majority of Y was consumed to form the protective oxide layer for the powder.

3.2.2 Heat treated specimens

Neither the grain size nor the texture was changed after the solution heat treatment, that was able to dissolve all the intermetallic precipitates (Fig. 7). Only the oxides were found in the matrix. Many of them were teared from the Mg matrix during polishing very likely because of the large difference in hardness between the oxides and the soft Mg matrix, leading to a population of elongated voids on the polished surfaces. However, these voids are an artefact of polishing and are not found within the struts.

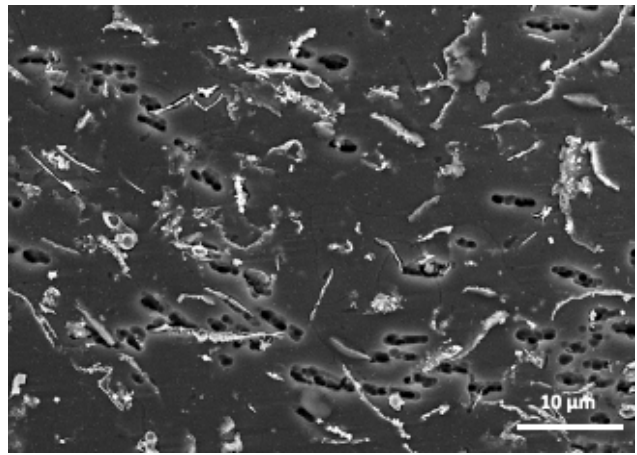

**Fig. 7.** SE image of WE43 Mg alloy in the T4 condition. The intermetallic precipitates are dissolved into the matrix. The voids correspond to oxide particles that were teared during polishing.

Artificial ageing at 250°C for 16 h (T6 treatment) resulted in two different types of precipitates, as shown in Fig. 8. Globular precipitates decorate the grain boundaries although can also be found occasionally within the grains. However, most of the precipitates have a plate shape and formed within grains. The plate-shaped precipitates grew parallel to the prismatic planes of the Mg hcp lattice, as shown in the HAADF image of a lamella parallel to the prismatic plane (Fig. 8a). They have a length of up to 500 nm and 80 - 200 nm in width, while the thickness was < 50 nm. The HAADF image of a lamella parallel to the basal plane (Fig. 8b) showed that these precipitates were sometimes interconnected to each other forming staircase structures between precipitates which grew parallel to different prismatic planes. The chemical analysis of the precipitates by EDS showed that they were rich in Nd, Gd and Y and depleted in Mg (Fig. 8d-g).

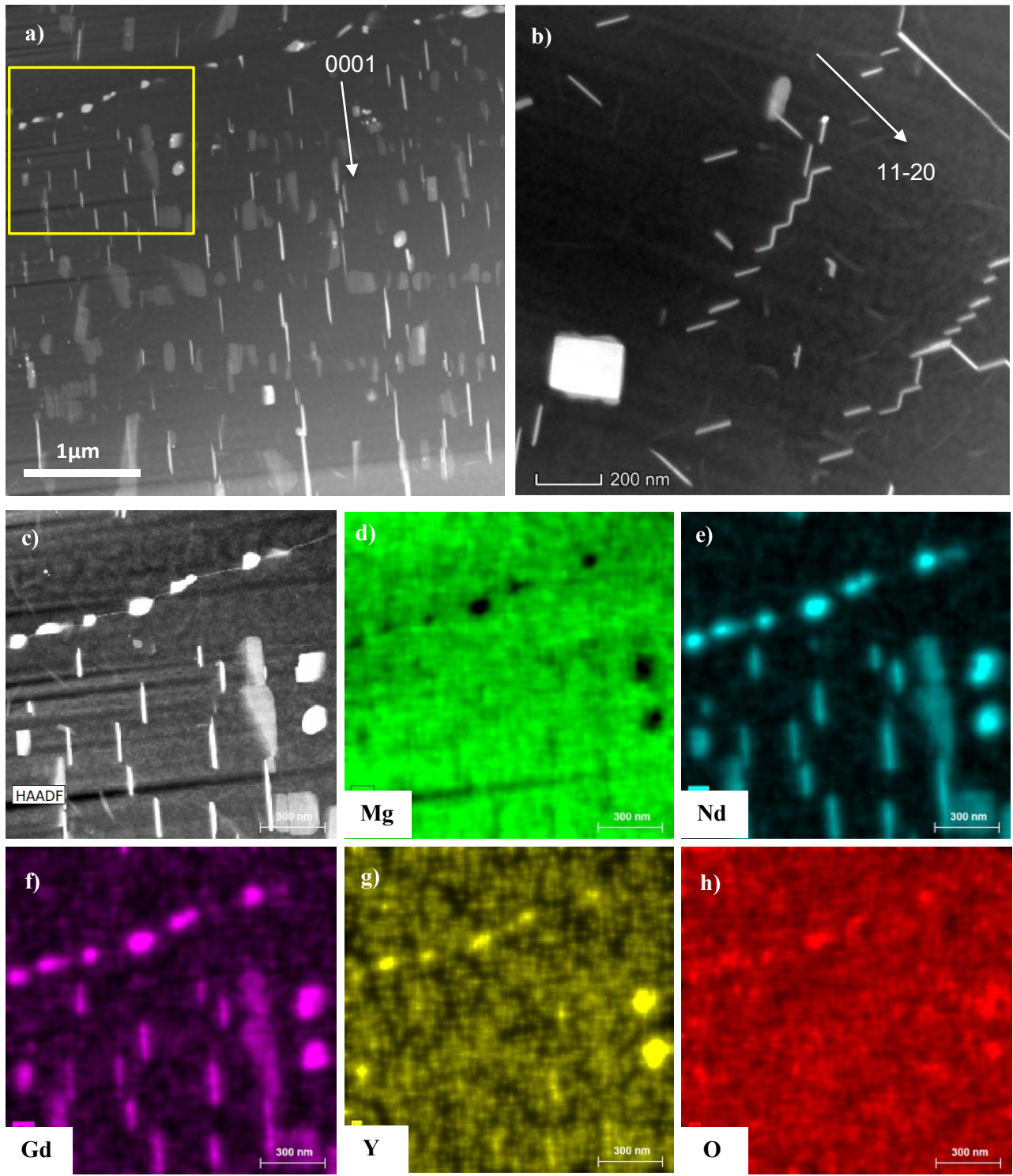

**Fig. 8.** HAADF images of the WE43 Mg alloy in the T6 condition. (a, b) Lamella parallel to the prismatic plane. (c) Lamella parallel to the basal plane. (d-h) EDS map of the different elements within the yellow box in (a).

The precipitates found in cast WE43 Mg alloys have been analysed in detailed in previous investigations [26-28]. These results indicate that the plate-shaped intermetallics parallel to the prismatic plane are $\beta_1$ precipitates with fcc structure and composition $Mg_3RE$, while the globular particles correspond to $\beta'$ precipitates with orthorhombic structure and composition $Mg_{12}YNd$.

*3.3 Corrosion behavior*

When immersed in DMEM, MS Mg scaffolds without PEO surface modification show relatively fast degradation rates. Both the corrosion rates and corrosion mechanisms depended on the metallurgical condition of the matrix. The cross-sections of the scaffolds obtained from the tomograms are plotted in Figs. 9a and b for the as-printed scaffold after 3 and 7 days in DMEM and for the scaffold in the T4 condition in Figs. 9c and d after the same time of immersion in DMEM. Corrosion progressed rapidly with immersion time in both cases and seemed to be localized in particular regions of the scaffold, where the struts and the lattice have disappeared, leading to large voids. However, struts in other regions of the scaffold did not seem to be affected by corrosion in the as-printed scaffold while corrosion was more homogeneous in the T4 condition. The cross-sections of scaffolds in the T6 condition after 2 and 3 days of immersion in DMEM are depicted in Figs. 9e and f, respectively. It is obvious that degradation took place at faster rates in this condition as indicated by the dimensions of the voids in the lattice as well as by the corrosion products, that appear as light grey regions. In fact, the scaffolds in the T6 condition lost their integrity after 2 days in DMEM.

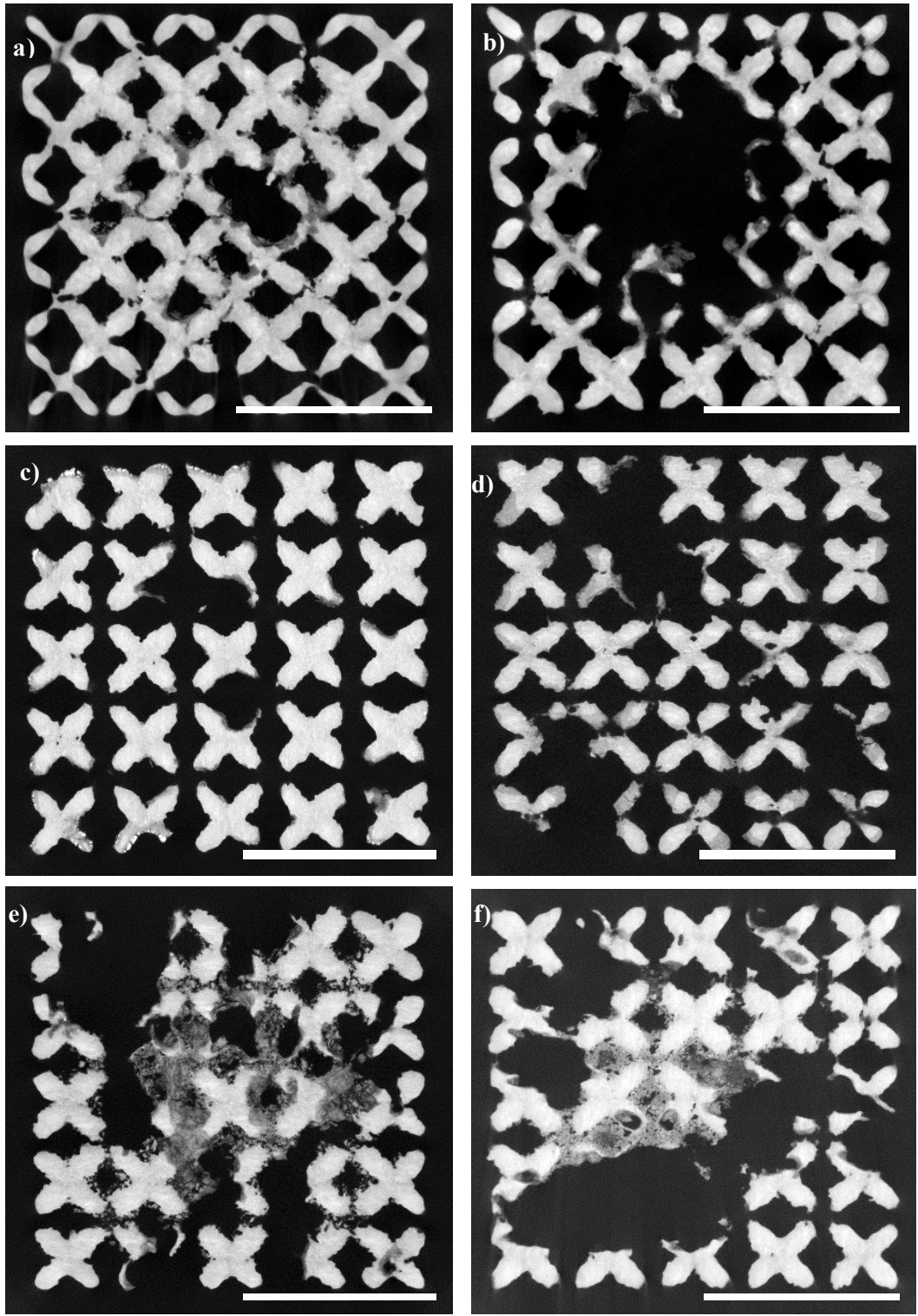

**Fig. 9.** XCT images of the central section of the scaffolds without PEO surface treatment (perpendicular to the building direction) after immersion in DMEM. (a) As-printed scaffold, 3 days. (b) As-printed scaffold, 7 days. (c) Scaffold in T4 condition, 3 days. (d) Scaffold in T4 condition, 7 days. (e) Scaffold in T6 condition, 2 days. (f) Scaffold in T6 condition, 3 days. The scale bars correspond to 5 mm.

The Mg mass loss, $\Delta M$, as a function of time was determined from the XCT images and from the volume of hydrogen generated during the process. In the former case, it was possible to distinguish the Mg metal in the lattice (white) from the corrosion products (grey) (Fig. 9) and the total volume of Mg remaining in the scaffold could be easily determined by binarizing the information in the 3D tomograms with a proper threshold. The results obtained from both methodologies are plotted in Fig. 10a and the corresponding corrosion rates, $CR$, calculated according to Eq. (1) are plotted in Fig. 10b. The mass losses and corrosion rates calculated from the tomograms are significantly higher than those determined from the hydrogen generation because they include full sections of the scaffold that broke off from the lattice even though not all the Mg has been corroded. Nevertheless, both strategies indicate - in agreement with the qualitative analysis of Fig. 9- that the corrosion of the scaffold in the T6 condition was much more severe, while the mass loss of scaffolds in the as-printed and T4 conditions was equivalent.

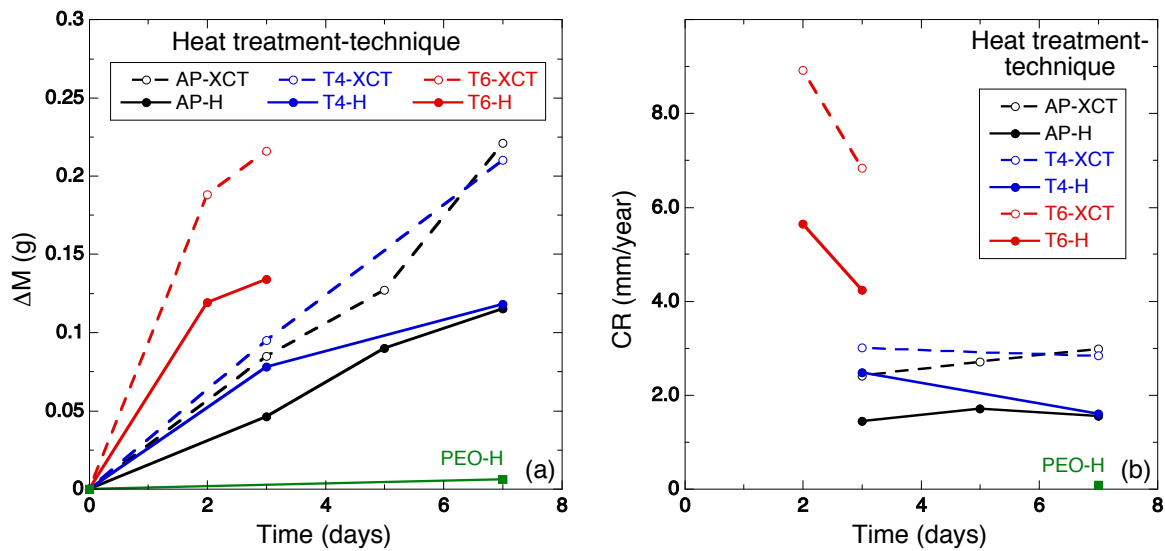

Fig. 10. (a) Mass loss, $\Delta M$, as a function of the immersion time in DMEM for MS WE43 Mg scaffolds processed by LPBF in different conditions: as-printed (AP) and subjected to T4 and T6 heat treatments. (b) Corrosion rates, $CR$, calculated from $\Delta M$ through eq. (1). Data obtained from XCT and hydrogen release are included in both figures. Moreover, the mass loss and corrosion rate measured by hydrogen release for a similar scaffold whose surface was modified by PEO is also plotted for comparison in both figures [21].

The mass loss and corrosion rate of an as-printed scaffold modified by PEO after 7 days immersed in DMEM are also plotted for comparison in Figs. 10a and 10b [21]. The differences between the scaffolds without and with surface treatment by PEO indicate that the microstructure of the Mg matrix plays secondary role -as compared with PEO surface modification- to control the degradation rate of Mg alloys.

*3.4 Mechanical properties*

3.4.1 Deformation of scaffolds with different strut size

The compressive nominal stress $S$ vs. engineering strain $\varepsilon$ curves of the scaffolds with different strut sizes (SS, MS, and LS) modified by PEO are plotted in Fig. 11. They present similar features, which are typical of porous structures deformed in compression. The initial linear elastic region was followed by a small non-linear zone which ended with a first maximum in strength. Further deformation led to successive minima and maxima in the load carried by the scaffold, but the maxima were fairly constant up to strains around 40%. An increase in strength due to densification is expected to be found at strains > 60% [31] but this regime was not attained in these tests.

The strength of the LS and MS scaffolds was significantly higher than that of the SS scaffolds and these differences should not be attributed to the density because they were still stronger when the nominal stress was divided by the relative density $\rho_r$ (calculated as 1 minus the porosity in Table 1). SE images of the MS scaffold deformed within the SEM (Fig. 11c-f) show that the first stress drop at an applied strain of 8% was associated with damage in the first row of cells by cracks parallel to the loading axis in the struts. The next maximum in stress was followed by the fracture of struts in the second row of cell when the applied strain reached 14%. This process continued by the progression of damage in successive rows, as shown in the SE images when the applied strain reached 30% and 54%.

The deformation of the LS scaffold can be seen in the video S4 in the Supplementary Material and show the same crashing process characterized by the fracture of successive rows of cells in the scaffolds, by cracks parallel to the compression direction in the struts. On the contrary, the deformation of the SS scaffold in the video S5 in the Supplementary Material show that the crashing process was only active at the beginning of the deformation and was replaced by the propagation of a shear band through the scaffold, oriented at 45º with respect to the loading axis, leading to a strong reduction in the load bearing capacity.

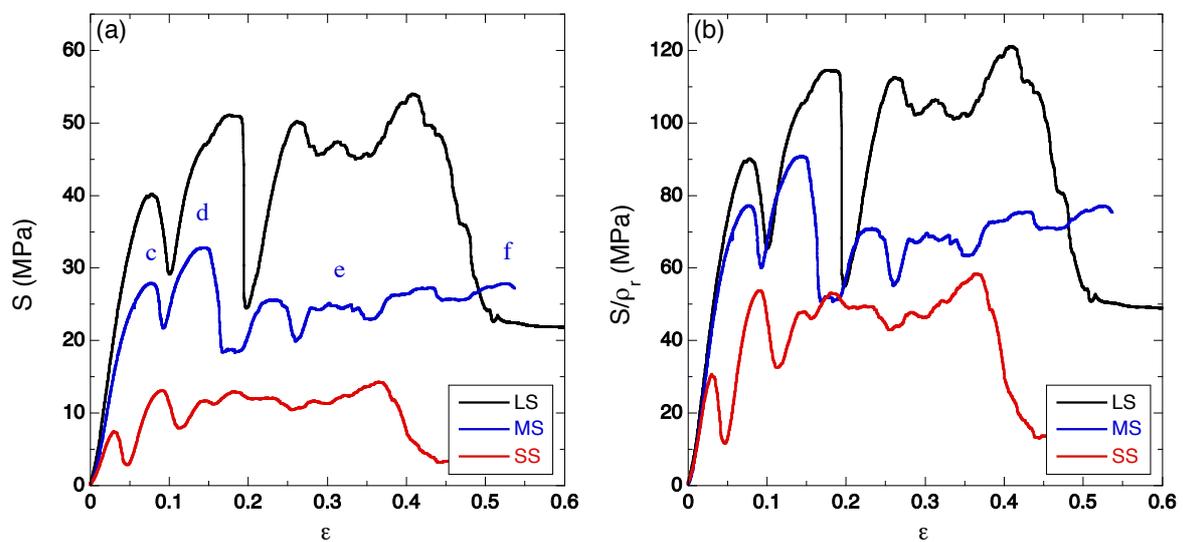

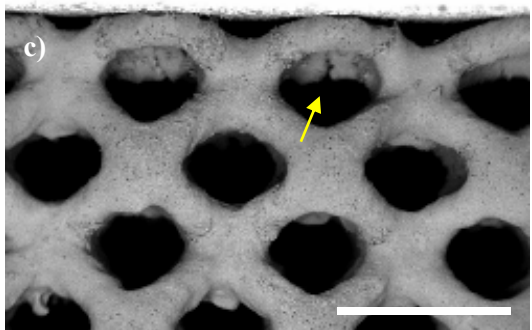
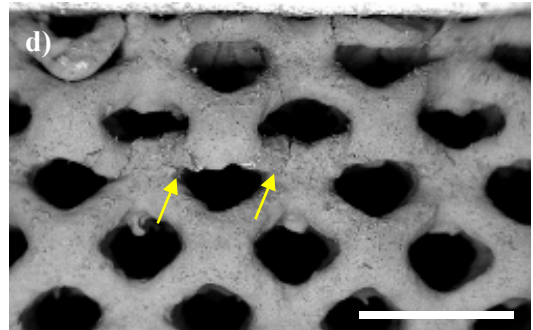
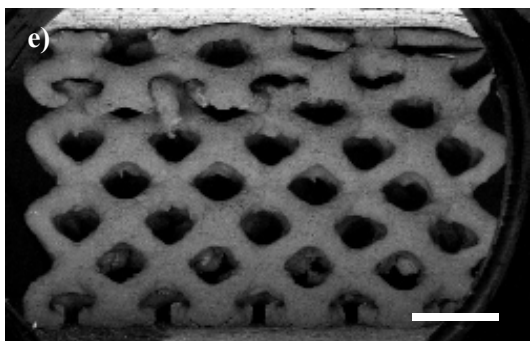
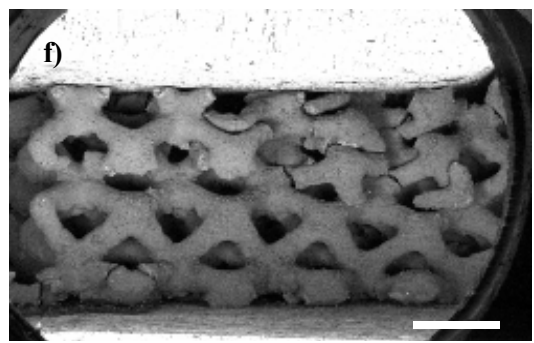

**Fig 11.** (a) Compressive nominal stress *S* vs. engineering strain ε curves of PEO-modified SS, MS and LS porous scaffolds deformed parallel to the building direction. (b) Idem as (a) but the nominal stress is normalized by the relative density of the scaffold, $\rho_r$. (c-f) SE micrographs of the MS scaffold deformed up to different strains marked in (a). Cracks in the struts parallel to the loading direction are indicated by the yellow arrows. The scale bars stand for 2 mm.

3.4.2 Mechanical anisotropy

The WE43 Mg scaffolds manufactured by LPBF showed a strong basal texture along the building direction. Thus, the Mg basal planes are perpendicular to the building direction and basal slip -which is the dominant deformation mechanism in Mg- is not favoured under compression parallel or perpendicular to the building direction. Instead, deformation by extension twinning is likely to dominate plastic deformation when the scaffolds are compressed in the direction normal to the building direction while compression along the building axis should be mainly accommodated by pyramidal slip. However, the critical resolved shear stress for extension twinning is much lower than that for pyramidal slip, which can potentially lead to a strong anisotropy on the mechanical response [32-34].

The compressive nominal stress *S* (divided by the relative density, $\rho_r$) *vs.* engineering strain $\varepsilon$ curves of the PEO-modified LS scaffolds deformed parallel (P) and normal (N) to the building direction are plotted in Fig. 12. Three different tests were carried in each orientation and no significant differences were found between both orientations up to 40% strain, indicating that it is unlikely that deformation by twinning took place in the struts. This extreme was confirmed by analysing by EBSD one cross-section of the scaffolds deformed in compression normal to the building direction after 20% strain. The pole figures indicated that extension twinning did not develop during deformation. It is known that the critical resolved shear stress for twining increases as the grain size decreases [35] and the average size of the largest grain population in the scaffolds (12 ± 6 μm) may be too small to allow twinning deformation. The load bearing capability of the scaffolds deformed perpendicular to the building direction increased for strains > 40%, very likely due to early densification in this orientation.

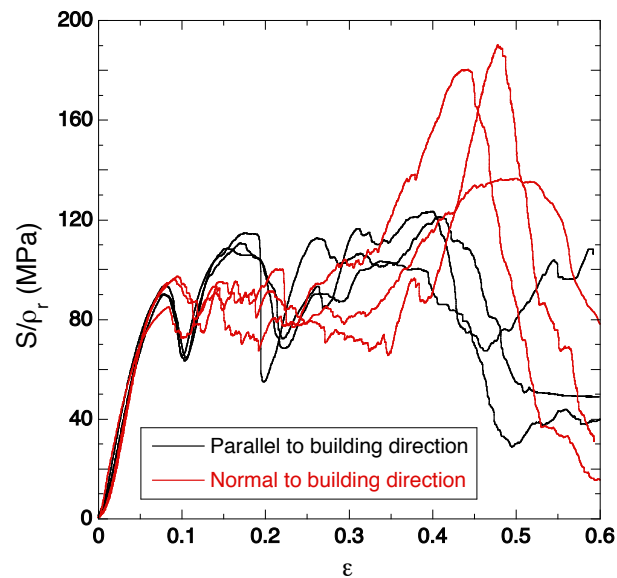

**Fig. 12.** Compressive nominal stress $S$ (divided by the relative density, $\rho_r$) *vs.* engineering strain $\varepsilon$ curves of as-printed WE43 Mg alloy LS scaffolds modified by PEO. Black lines stand for specimens tested parallel to the building direction and red lines for the ones tested in the direction normal to the building direction.

3.4.3 Effect of heat treatments

The effect of the heat treatments on the mechanical properties was analysed in the case of MS scaffolds which were not modified by PEO. The compressive nominal stress $S$ *vs.* engineering strain $\varepsilon$ curves of three MS scaffolds in the as-printed, T4 and T6 conditions are plotted in Fig. 13. They show that the heat treatment did not modify the elastic modulus, yield strength and the strength of the scaffolds. The only significant differences were found for $\varepsilon > 40\%$ in the heat treated scaffolds, which showed a large increase in the load bearing capacity beyond this strain, while the strength of the as-printed scaffold decreased up to $\varepsilon = 40\%$. The video S6 in the Supplementary Material shows the mechanisms of deformation and fracture of the MS scaffold in the T4 condition. Collapse of the scaffolds occurs layer by layer by plastic deformation of the struts and without any lateral expansion. This behaviour is different from the one in Videos S4 and S5 of as-printed scaffolds where compression by crushing is associated to the fracture of the scaffolds and to lateral expansion. As a result, densification in the as-printed scaffolds begins at ε > 40%.

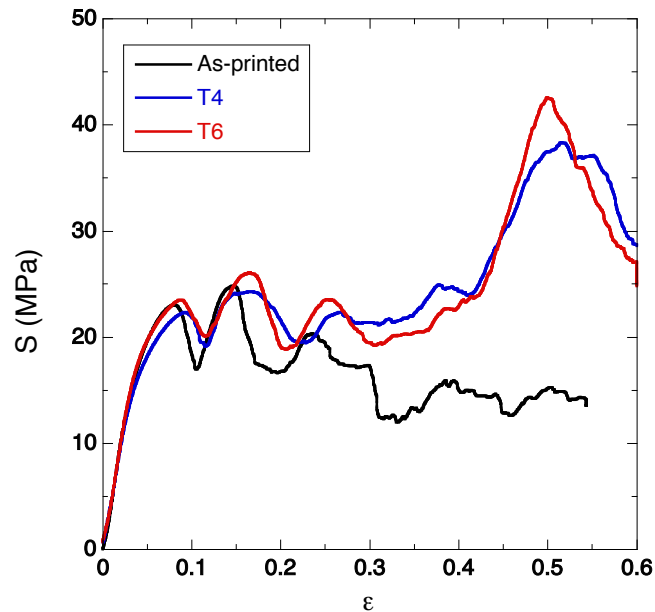

**Fig. 13.** Effect of heat treatment on the compressive nominal stress *S vs.* engineering strain ε curves of WE43 Mg alloys MS scaffolds without surface modification.

3.4.4 Mechanical properties after corrosion

The compressive *S vs.* ε curves after different times immersed in DMEM solution are plotted in Figs. 14a, b and c for MS scaffolds in the as-printed, T4 and T6 conditions, respectively. The rapid degradation of the mechanical properties with immersion time is obvious in all cases and led to reductions in the elastic modulus and in the strength of the scaffold. Degradation was particularly fast in the scaffold in the T6 condition, whose strength was below 10 MPa after two days of immersion in DMEM. The degradation of the mechanical properties of the scaffolds in the T4 and as-printed conditions was similar, in agreement with the corrosion rates in Fig. 10b.

Corrosion lead in all cases to a change in the shape of the *S-ε* curves. The curves of the scaffolds before immersion presented several peaks and valleys, which were associated with the progressive failure of different layers of the ordered structure of the scaffold. However, the symmetry of the scaffold is destroyed by corrosion and failure no longer takes place in an orderly fashion. As a result, the initial yield strength decreases dramatically during corrosion and the stress carried by the scaffold remains constant with the applied strain.

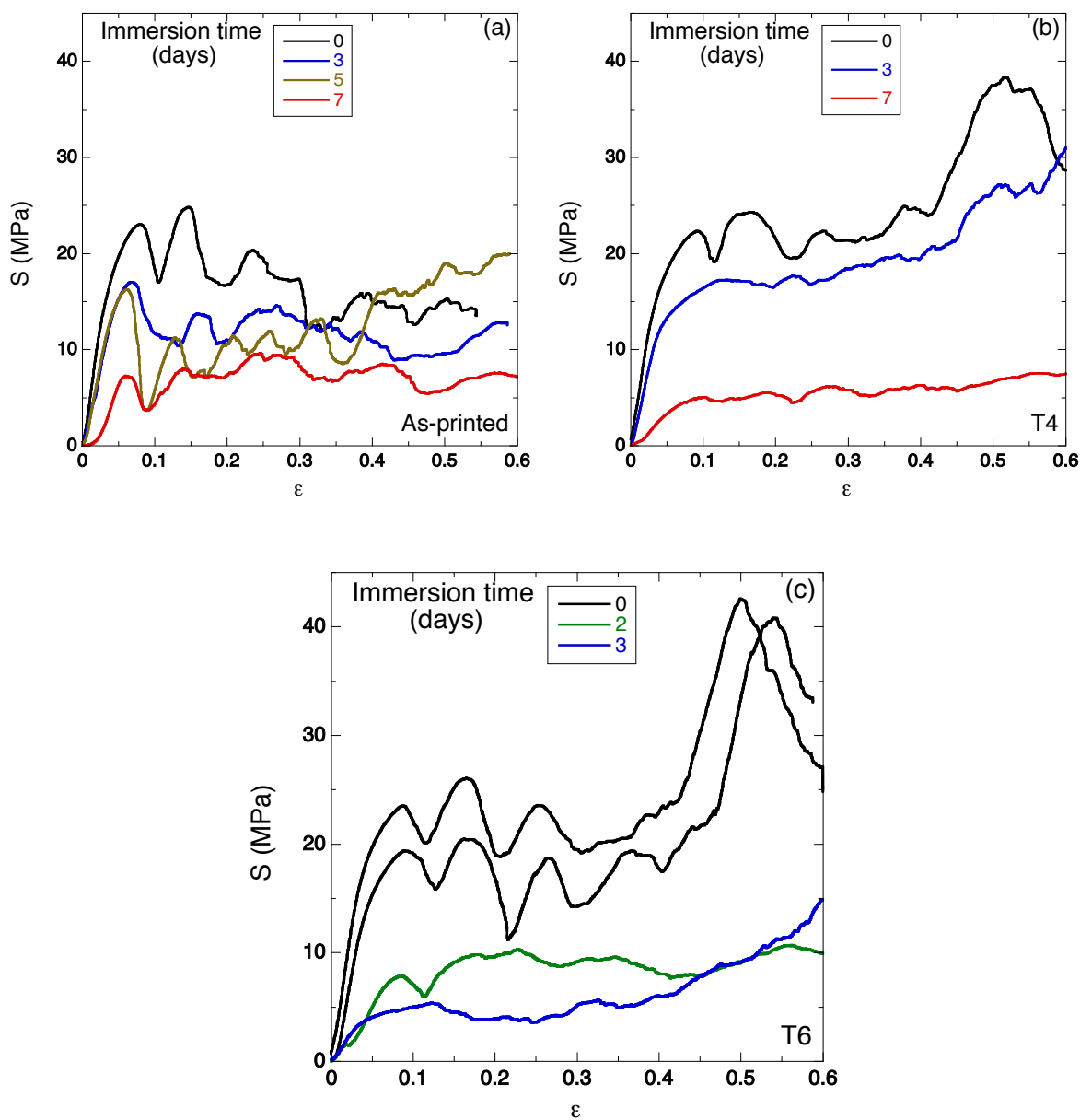

**Fig. 14.** Effect of immersion in DMEM solution on the compressive nominal stress *S vs.* engineering strain *ε* curves of WE43 Mg alloys MS scaffolds without PEO surface modification. (a) As-printed. (b) T4 condition. (c) T6 condition.

*3.5 In vitro cytotoxicity*

Different cell culture-based tests were conducted in the MS WE43 Mg scaffolds with and without PEO surface modification in the as-printed condition. Indirect tests were conducted on the extracts derived from the different groups, as shown in Fig. 15, to evaluate the effects of the remaining components of the dissolved testing material, such as debris, impurities, or substances from the base material and surfaces.

Significant differences between the negative control and both testing groups in comparison to the positive control group were present within the LDH test assay. With equivalent (WE43) and even diminished absorbance values (WE43 PEO), possibly due to assay interference, both experimental groups exhibited no signs of direct cell damage. This would present itself as spectrophotometrically measurable highly colored formazan, which is produced by the release of soluble cytosolic enzyme into the cell culture medium due to cell death and damage of the plasma membrane, respectively.

With regard to cellular metabolic activity, both the XTT assay and BrdU assay showed equivalently elevated absorbance values for WE43 PEO specimens, therefore indicating cell proliferation. XTT quantifies cellular metabolic activity via colorimetric assay of added, extracellular tetrazolium salts, which are transformed into colored formazan derivatives via mitochondrial dehydrogenase, while BrDU detects the quantity of 5-bromo-2'-deoxyuridine (BrdU) incorporated into cellular DNA during cell proliferation using an anti-BrdU antibody. While significantly different to the toxic positive control, both testing groups showed reduced activity when compared to the non-toxic negative control. In contrast to direct cell damage assessed by LDH, WE43 scaffolds without PEO surface modification showed significantly decreased metabolic activity in the XTT and BrdU assays, hereby indicating cytotoxic potential.

However, WE43 scaffolds whose surface was modifed by PEO showed a significant increase in metabolic activity and cellular proliferation for both tests with means slightly exceeding or falling below the 70% threshold defined by the ISO 10993-5:2018 standard. Hence, cytotoxic potential based on indirect testing of extracts is not present after PEO surface modification is applied to the tested groups.

In order to obtain a more complete picture of the biological interaction of WE43 Mg scaffolds, complementary direct testing by live dead stains was performed. In this study, live dead stains were performed using two different solid materials as references (RMA and TCC). Fig. 16 shows the overlay of green and red emitting stains in different magnifications as well as non-stained cells for test verification (right column). While no viable green cells can be observed, only a few rounded and dead red stained cells can be found on the toxic RMA control. In contrast, the negative reference material shows no dead, but only a plethora of vital spindle-shaped and green fluorescence emitting cells on the non-toxic TCC surface, thus validating the liability and specific range of the conducted experiments. With regards to the tested scaffold groups, it becomes obvious that there is a visible difference in cell attachment and vitality as a function of the surface state. Almost no cells bound to the surface at all (only minor rounded and dead cells) can be identified in the scaffolds without PEO surface modification. On the contrary, the scaffolds whose surface was modified by PEO show a large number of vital green cells with no visible difference in number when compared to the non-toxic negative control. While the appearance of pseudopodia and associated cell attachment can be scarcely assessed on the negative control, it can only be assumed within the PEO group and was therefore neglected in terms of further evaluation. In summary, WE43 Mg scaffolds surface treated by PEO display acceptable cell interaction with low cytotoxic potential expressed via indirect and direct cell culture testing. Results always approximated the reference values of the negative controls.

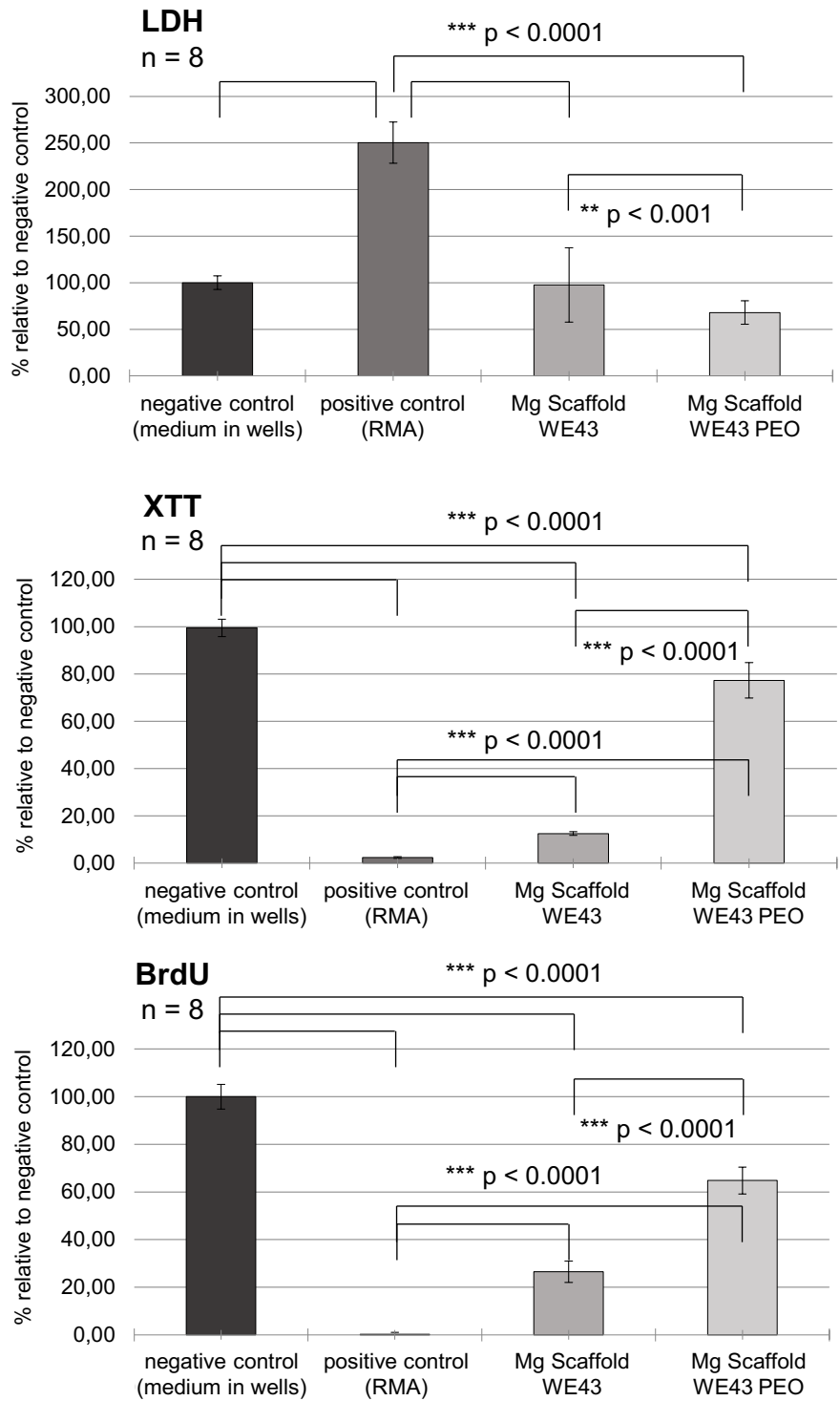

**Fig. 15.** Results of extract-based cell culture assays on WE43 Mg alloy scaffolds with and without surface modification by PEO.

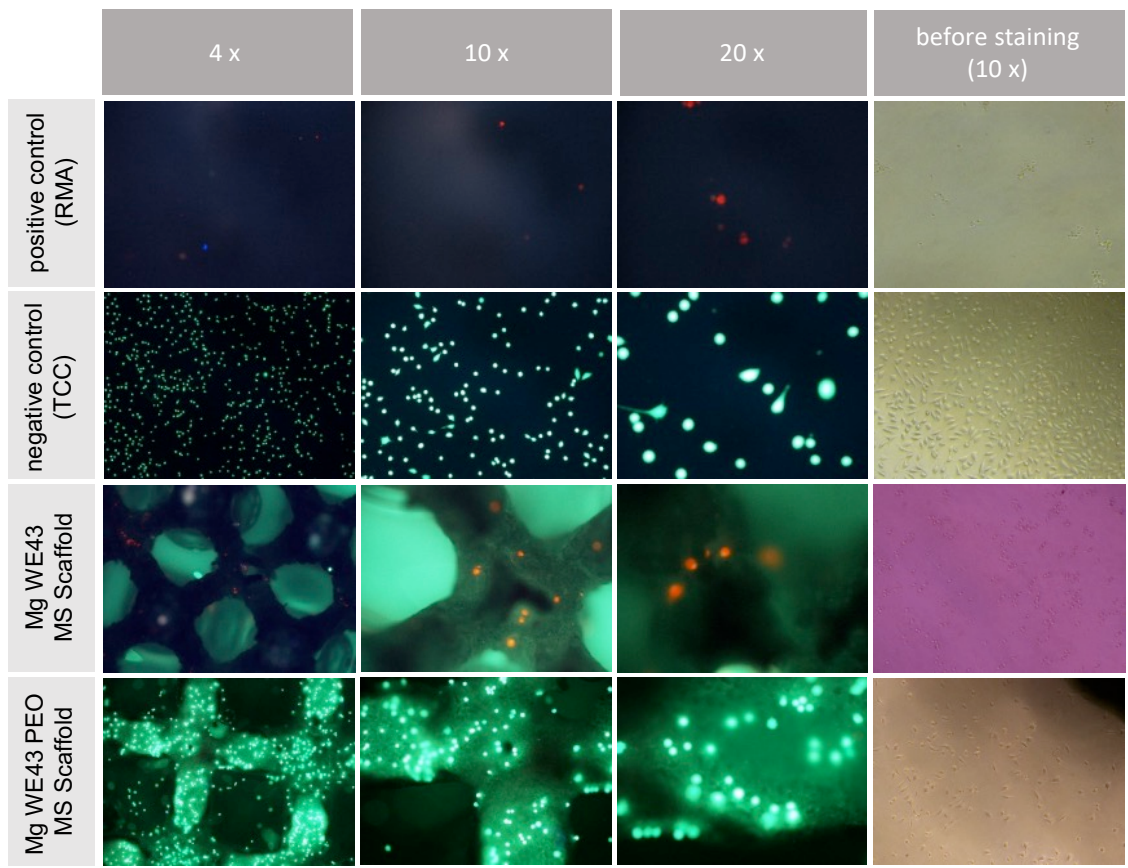

**Fig. 16.** Live dead stains of L929 mouse fibroblasts on WE43 Mg alloy scaffolds with and without surface modification by PEO.

## 4. Discussion

*4.1 Microstructure*

The WE43 Mg alloy scaffolds show the typical microstructure of alloys manufactured by LPBF with clear traces of the layer by layer process under optical microscopy. EBSD results showed a bi-modal grain size distribution: small grains with random texture were found near the melt pool boundary while large grains had a strong basal texture parallel to the building direction (Fig. 3). Bär *et al.* [23] analyzed the 'last melt pool' of a WE43 Mg alloy block processed by LPBF which is not affected by following depositions and found that it contains elongated grains originating from the melt pool boundary and equiaxed grains at the center with random texture. But the grain morphology below the last melt pool was very similar to the one depicted in Fig. 3. Therefore, the large grains with basal texture develop during the re-melting of each layer as a result of the printing of the successive layers.

In addition to the pre-existing Y-rich oxide particles, three precipitate structures are found throughout the specimens at, above and below the melt pool boundaries. Similar structures have been previously observed in alloys processed by rapid solidification. For instance, Gianoglio *et al.* [36] found both wavy and broken lines formed by precipitates parallel to the heat flow direction, similar to those in Figs. 5b and c, in a laser-melted Al-3Er alloy, that were denominated fibrous and uncoupled eutectic structures, respectively. They argued that the columnar crystals of the primary phase grow epitaxially when solidification starts, and eutectic phases appear in between because of lateral rejection of solute.

Compared to the vertical structures of precipitates, the horizontal lines show lower fluctuations and wider spacing. They consist of either lines of round nano-precipitates or consecutive lines with nano-porosities (Fig. 5d and e). These horizontal lines have been referred to as 'banded structures' [37-42]. Mckeown *et al.* [37] observed similar banded structures at the melt pool center in an Al-4Cu alloy solidified during *in situ* TEM imaging. They were formed by alternating bands of microsegregation-free, supersaturated solid solution α-Al and bands of non-equilibrium eutectic phases. These structures develop when the solidification front reaches a critical velocity that leads to an interfacial instability [43] and the solidification front velocity exceeds the solute diffusivity in the liquid, leading to a rapid increase of the velocity-dependent partition coefficient toward unity [44]. Thus, the formation of these banded structures can be explained as follows. When a melt pool is formed, the movement of the liquid-solid interface is slow at the melt pool boundary and, therefore, the precipitates are large. As the solidification front accelerates, Mg crystals grow along the heat flow direction (parallel to the building direction) and the eutectic phase forms in the form of wavy or broken lines due to lateral rejection of solute atoms (Nd, Gd and Y). As the solid-liquid interface continues accelerating, some solute atoms are trapped in the Mg crystals, while some are rejected in front of the moving interface. Horizontal lines of precipitates are formed from this supersaturated Mg solid solution when the Gibbs free energy for nucleation is low enough. The solid-liquid interface is depleted from solute after the formation of the horizontal line of precipitates and the solute concentration perpendicular to the horizontal lines shows a gradient near each line of precipitates (Fig. 6c). This process is repeated periodically along the heat flow direction, resulting in a relatively even horizontal line spacing.

Grain size and texture of the scaffolds did not change during the heat treatments, but the precipitate structure was dramatically changed by the heat treatments. The solution heat treatment at 525 ºC dissolved all the intermetallic precipitates in the Mg matrix which was only decorated by the oxide particles. Ageing at 250 ºC during 16 hours leads to the formation of β' and $β_1$, as in cast WE43 Mg alloys [26-28]. Other types of precipitates, such as β'' and β, were not found. β'' is only present at very early stages of ageing or at lower temperatures and the equilibrium β phase results from an *in situ* transformation of $β_1$ after 24 hours of ageing at 250 °C [26-28].

*4.2 Corrosion behaviour*

In an aqueous environment, Mg and its alloys react following the equations [45]:

$$Mg_{(s)} \rightarrow Mg^{2+}_{(aq)} + 2e^- \tag{2}$$

$$2H_2O_{(aq)} + 2e^- \rightarrow H_{2(g)} \uparrow + 2OH^-_{(aq)} \tag{3}$$

$$Mg^{2+}_{(aq)} + 2OH^-_{(aq)} \rightarrow Mg(OH)_{2(s)} \downarrow \tag{4}$$

The $Mg(OH)_2$ deposits on the Mg matrix and the corrosion product layer usually contains an inner MgO layer and an external $Mg(OH)_2$ layer [46]. The inner MgO layer is not dense enough to provide effective protection against corrosion, while the external $Mg(OH)_2$ layer can react with the chloride ions in physiological conditions to form soluble $MgCl_2$ according to

$$Mg(OH)_{2(s)} + 2Cl^-_{(aq)} \rightarrow MgCl_{2(aq)} + 2OH^- \tag{5}$$

Thus, the protective layer is broken leading to high corrosion rates. In addition, the presence of impurities and second phases contributes to the accelerated corrosion of Mg alloys because of the development of galvanic corrosion as secondary phases usually have a more positive corrosion potential than that of α-Mg.

The corrosion rates of the MS scaffolds, depicted in Fig. 10b were very high in all cases and exceeded by a large amount the maximum corrosion rates of 0.5 mm/year that are required for biodegradable implants [47]. Nevertheless, this threshold can be easily achieved if the scaffolds are modified by PEO, leading to corrosion rates < 0.1 mm/year [21]. The results in Fig. 10 also show that the corrosion resistance of the scaffolds without surface modification by PEO is highly dependent on the heat treatment. The corrosion rates in the T6 condition were higher (by a factor of 2 to 3) than those found in as-printed and T4 conditions, which showed similar rates (slightly higher in the T4 condition). The heat treatment also influenced the progression of corrosion. It is more localized in the as-printed and T6 conditions (Figs. 9b and 9f, respectively) while more homogeneous corrosion of the struts is observed in the T4 condition (Fig. 9c). The differences in the localization of the corrosion between the as-printed and T4 conditions can be observed in the videos S7 and S8, respectively, in the supplementary material, which show successive cross-sections of the tomogram of the scaffolds after 7 days of immersion in DMEM. The highest corrosion rate as well as the localization of corrosion in the T6 condition can be easily attributed to galvanic corrosion triggered by the large precipitates formed during high temperature ageing (Fig. 8). In particular, the globular β' precipitates along the grain boundaries are likely to accelerate localized corrosion at the grain boundaries. All the intermetallic precipitates were dissolved in the matrix in the T4 condition, leading to a large improvement in the corrosion resistance and to a more homogeneous degradation of the scaffold.

The lowest corrosion rate was found in the as-printed scaffolds and this result was surprising. Although no clear explanation for this behaviour was found, several reasons can be responsible for this behaviour. The intermetallic precipitates in the as-printed condition are much smaller than those in the T6 condition (compare Figs. 5 and 8) because of the extremely high cooling rates in the former and the reduction in precipitate size can alleviate the micro-galvanic effect [14]. Moreover, the precipitates in the as-printed alloy are far from equilibrium and they might not be favourable to promote galvanic corrosion. For instance, Shuai et al. [48] reported that the potential difference between $Mg_{12}Nd$ and Mg is relatively small because the standard electrode potential of Nd (-2.43 V [49]) is lower than that of Mg and these precipitates will not contribute significantly to galvanic corrosion. Similar behaviours could be expected from other small non-equilibrium precipitates in the as-printed condition.

*4.3 Mechanical properties*

The main parameters that define the mechanical behaviour of the scaffolds according to the definitions in ISO 13314:2011 have been plotted as a function of the strut diameter in Fig. 15 for PEO-modified scaffolds. They are the elastic modulus, *E*, the first maximum strength in the *S*-ε curves, $S_1$, the plateau strength (average stress carried by the scaffold between 20% and 30% strain), $S_p$, and the energy absorption (area below the *S*-ε up to 50% strain), *W*. All of them increase rapidly with the strut diameter and those of the LS scaffolds (with a strut diameter of 793 ± 39 μm) show an elastic modulus of ≈ 0.8 GPa, which is in the range of the trabecular bone (0.5-20 GPa [50]). Nevertheless, strength of the scaffolds is significantly lower than that of cortical bones (90-205 MPa [51-52]) and these results indicate that scaffolds with thicker struts and/or different lattice design should be designed to improve the mechanical properties of the scaffolds for biomedical applications, at least if demanding the replacement of cortical structures under significant load. Thicker struts will also improve the corrosion resistance of the lattices because they reduce the overall surface of the scaffold.

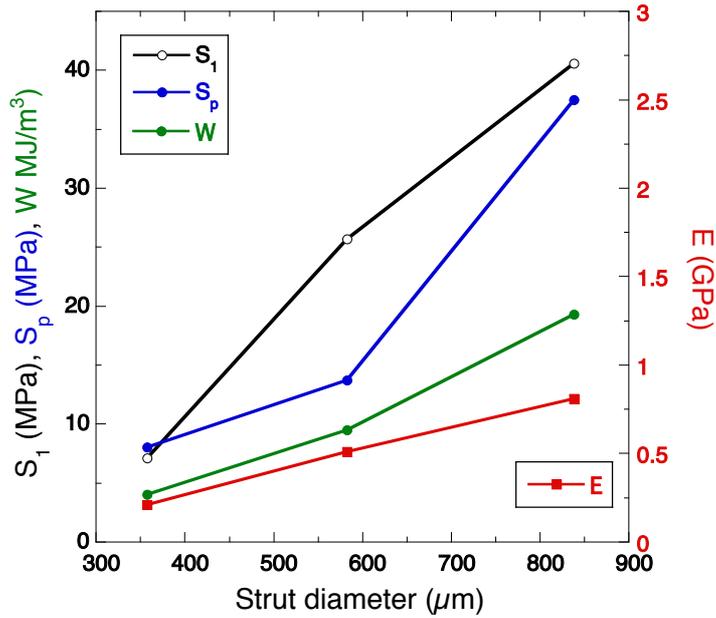

**Fig. 17.** Mechanical properties of the PEO-modified WE43 Mg scaffolds along the building direction as a function of the strut diameter.

The effect of the heat treatments on the mechanical properties of the scaffolds (Fig. 13) was limited. No large differences were found in the elastic modulus, the first maximum strength and the plateau strength even though no intermetallic precipitates were present in the T4 condition and the precipitate distributions were very different between the as-printed and T6 conditions. These results indicate that the small grain size and the dispersion of the oxide particles controlled the strength of the WE43 Mg alloy processed by LPBF, and these parameters were not modified by the heat treatments. The only relevant effect of the heat treatments was to modify the failure mode of the struts from brittle (by cracks parallel to the loading direction) in the as-printed condition to ductile (by plastic collapse of the struts) in the heat treated scaffolds. As a result, densification took place at lower strains in the heat treated scaffolds.

Overall, these results indicate that the mechanical properties of LPBF magnesium scaffolds should be improved to reach the strength of cortical bone for bone implant applications. This latter objective can be achieved through the optimization of the mechanical properties of the Mg alloys [53] and the design of lattice structures with improved strength [54].

*4.4 In vitro cell response*

Indirect extract based assays (LDH, XTT and BrdU), as well as direct cell seeding and staining were performed as a part of this study in order to assess the interaction of WE43 Mg alloy scaffolds with basic cellular activity. The negative (blank medium for indirect and TCC for direct testing) and positive controls (RMA for direct and indirect testing) demonstrated the anticipated toxic and non-toxic reactions, thus validating the experimental procedure. Based on potential adverse effects caused by the release of substances during extract formation, both tested Mg scaffold groups showed no direct toxicity by means of LDH. Further assessment with regards to the metabolic state and proliferation of cells (XTT, BrdU) revealed, that the scaffolds whose surface was modified by PEO expressed significantly increased absorbance approximating the standardized 70% threshold, while the non modified group rather failed the testing. These results lie well within expectation and can be confirmed by previous works concerning the in-vitro assessment of Mg materials [55]. While there seems to be no cytotoxic potential deriving from RE-based magnesium alloys themselves, the vast reaction of the bare metal surface leads to high values of initial hydrogen gas evolution and local pH shifts, thus impairing cell metabolism. While adverse effects in monocultural cell procedures are visible, past in-vivo studies did not demonstrate any immediate cytotoxicity of Mg-based devices.

Similar observations were found for direct live/dead staining. While no viable cells were observed on the Mg scaffold without surface modification by PEO, only few dead cells were identified at all. The surface therefore seems to repel cell attachment, which is in line with the reported hydrogen gas evolution, hereby discouraging cells to adhere to the surface. In contrast, the WE43 Mg alloy scaffolds whose surface was modified by PEO showed adherent cell monolayers, very much comparable to the non-toxic negative control. The passivating ceramic-like surface not only seems to offer a suitable niche for adhering cells, but also throttled the release of degradation by-products and thus encouraged hardly any signs of cell impairment.

## 5. Conclusions

Porous scaffolds of WE43 Mg alloy with a body-centered cubic cell pattern were manufactured by LPBF with different strut diameters in the range 250 μm to 750 μm. The microstructure was formed by a bimodal grain size distribution with large grains (12 ± 6.1 μm) and small equiaxed grains (2.1 ± 1.0 μm). The former presented a strong basal texture while the latter were randomly oriented. Irregular oxide particles containing Y were homogeneously distributed throughout the microstructure. Intermetallic precipitate rich in Nd, Gd and Y were found in the as-printed scaffolds forming three different types. In particular, banded structures parallel and perpendicular to the melt pool were formed as a result of segregation of the solute atoms during rapid solidification. The precipitate structures disappeared after a solution heat treatment at 525 °C for 8 h and globular β' and plate-shape $β_1$ precipitates were formed at the grain boundaries and within the grains, respectively, after ageing at 250 ºC for 16 h.

Scaffolds without PEO surface treatment showed increased corrosion rates in DMEM, which exceeded the suggested corrosion rates of 500 μm/year required for biodegradable implants. The maximum corrosion rates were found in the scaffolds aged at 250 ºC and were attributed to the galvanic corrosion induced by the intermetallic precipitates. The best corrosion rates were measured in the as-printed and solution treated scaffolds and were attributed to the small size of the precipitates in the former and to the lack of intermetallic precipitates in the latter. Nevertheless, surface treatments, such as PEO, significantly improve corrosion rates.

The mechanical properties of the scaffold increased with the strut diameter due to the lower porosity and a change in the failure mechanisms. Scaffolds with an average strut diameter of 250 μm failed by the propagation of shear band at 45º with respect to the compression axis while those with thicker struts showed progressive failure of successive rows of cells in the scaffold. Heat treatments did not modify the mechanical properties, but the failure mechanisms of the strut changed from brittle failure to plastic collapse in the heat treated conditions.

The elastic modulus and strength in compression of the scaffolds with the largest strut diameter were around 0.8 GPa and around 40 MPa, respectively, being within the range of spongy bone, but diminished when compared to those of cortical bone. Moreover, the modulus decreased quickly after immersion in DMEM and the degradation in mechanical properties followed the same trends observed in the corrosion rates. The corrosion resistance could meet the requirements of biodegradable metal implants by means of PEO surface treatments while the mechanical properties of the porous scaffolds can be improved to reach the strength of cortical bone by using different lattice designs and stronger Mg alloys.

Biocompatibility tests demonstrated that WE43 Mg alloy scaffolds (with high surface to volume ratio) without PEO surface treatment expressed cytotoxic potential to a certain extent, namely when subjected directly to cells or culture media. In contrast, if the scaffolds were surface-treated by PEO, they exhibited ameliorated compatibility with cells by providing a favorable surface morphology for adherence and limited release of degradation by-products. Accordingly, cell culture testing by direct or indirect exposure rendered satisfactory results with no or only limited cytotoxic potential. Hence, applying suitable surface modifications seems mandatory when designing and manufacturing Mg scaffolds with large surface areas.

**Acknowledgements**

This investigation was supported by the European Research Council (ERC) under the European Union's Horizon 2020 research and innovation programme (Advanced Grant VIRMETAL, grant agreement No. 669141). Additional support was received from the European Union's Horizon 2020 research and innovation programme under the Marie Skłodowska-Curie grant agreement No 813869, from the Spanish Government through grant PID2019-109962RB-100, and from the Madrid regional government, under programme S2018/NMT-4381-MAT4.0-CM. Technical assistance from Manuel Avella, Jose Sanchez and Javier Garcia is gratefully acknowledged.

# References

[1] Roddy E. et. al, Eur J Orthop Surg Traumatol. 2018 Apr; 28(3):351-362.

[2] N. M. Haines, W. D. Lack, R. B. Seymour, and M. J. Bosse, "Defining the Lower Limit of a 'Critical Bone Defect' in Open Diaphyseal Tibial Fractures.," *J. Orthop. Trauma*, vol. 30, no. 5, pp. e158-63, May 2016, doi: 10.1097/BOT.0000000000000531.

[3] Stewart S. et. al, Chapter 24 - Bone Regeneration. Translational Regenerative Medicine. 2015, Pages 313-333.

[4] Ibrahim A., 13 - 3D bioprinting bone. 3D Bioprinting for Reconstructive Surgery Techniques and Applications. 2018, Pages 245-275.

[5] M. Moravej and D. Mantovani, "Biodegradable Metals for Cardiovascular Stent Application: Interests and New Opportunities," *International Journal of Molecular Sciences*, vol. 12, no. 7. 2011, doi: 10.3390/ijms12074250.

[6] Y. Liu *et al.*, "Fundamental Theory of Biodegradable Metals—Definition, Criteria, and Design," *Adv. Funct. Mater.*, vol. 29, no. 18, 2019, doi: 10.1002/adfm.201805402.

[7] M. B. M.G. Laires, C.P. Monteiro, "Role of cellular magnesium in health and human disease," *Front. Biosci.*, vol. 9, no. 9, p. 262, 2004.

[8] N. Wu and A. Veillette, "Magnesium in a signalling role," *Nature*, vol. 475, no. 7357, pp. 462–463, 2011, doi: 10.1038/475462a.

[9] R. J. B. J.H. De Baaij, J.G. Hoenderop, "Magnesium in man: implications for health and disease," *Physiol. Rev.*, vol. 95, no. 1, pp. 1–46, 2015.

[10] M. P. Staiger, A. M. Pietak, J. Huadmai, and G. Dias, "Magnesium and its alloys as orthopedic biomaterials: A review," *Biomaterials*, vol. 27, no. 9, pp. 1728–1734, 2006, doi: https://doi.org/10.1016/j.biomaterials.2005.10.003.

[12] H. Q. Ang, T. B. Abbott, S. Zhu, C. Gu, and M. A. Easton, "Proof stress measurement of die-cast magnesium alloys," *Mater. Des.*, vol. 112, pp. 402–409, 2016, doi: https://doi.org/10.1016/j.matdes.2016.09.088.

[13] D. Zhao, F. Witte, F. Lu, J. Wang, J. Li, and L. Qin, "Current status on clinical applications of magnesium-based orthopaedic implants: A review from clinical translational perspective," *Biomaterials*, vol. 112, pp. 287–302, 2017, doi: https://doi.org/10.1016/j.biomaterials.2016.10.017.

[14] J.-W. Lee *et al.*, "Long-term clinical study and multiscale analysis of in vivo biodegradation mechanism of Mg alloy," *Proc. Natl. Acad. Sci.*, vol. 113, no. 3, pp. 716 LP – 721, Jan. 2016, doi: 10.1073/pnas.1518238113.

[15] G. Ryan, A. Pandit, D.P. Apatsidis, Fabrication methods of porous metals for use in orthopaedic applications, Biomaterials 27 (13) (2006) 2651-2670.

[16] M.P. Staiger, I. Kolbeinsson, N.T. Kirkland, T. Nguyen, G. Dias, T.B.F. Woodfield, Synthesis of topologically-ordered open-cell porous magnesium, Mater. Lett. 64 (23) (2010) 2572-2574.

[17] D. Carluccio, A. G. Demir, M. J. Bermingham, M. S. Dargusch, Challenges and Opportunities in the Selective Laser Melting of Biodegradable Metals for Load-Bearing Bone Scaffold Applications. Metallurgical and Materials Transactions A 51, 3311- 3334, 2020, doi: 10.1007/s11661-020-05796-z

[18] L. Jauer, B. Jülich, M. Voshage, W. Meiners, "Selective laser melting of magnesium alloys,"

*Eur. Cells Mater.*, vol. 30, p. 1, 2015.

[19]  Y. Li *et al.*, "Additively manufactured biodegradable porous magnesium," *Acta Biomater.*, vol. 67, pp. 378–392, 2018, doi: 10.1016/j.actbio.2017.12.008.

[20]  Y. Li *et al.*, "Biodegradation-affected fatigue behavior of additively manufactured porous magnesium," *Addit. Manuf.*, vol. 28, no. December 2018, pp. 299–311, 2019, doi: 10.1016/j.addma.2019.05.013.

[21]  A. Kopp *et al.*, "Influence of design and postprocessing parameters on the degradation behavior and mechanical properties of additively manufactured magnesium scaffolds," *Acta Biomater.*, vol. 98, pp. 23–35, 2019, doi: 10.1016/j.actbio.2019.04.012.

[22]  M. Echeverry-Rendon, V. Duque, . Quintero, M. C. Harmsen, F. Echeverria, Novel coatings obtained by plasma electrolytic oxidation to improve the corrosion resistance of magnesium-based biodegradable implants. Surface & Coatings Technology 354, 28-37, 2018.

[23]  F. Bär *et al.*, "Laser additive manufacturing of biodegradable magnesium alloy WE43: A detailed microstructure analysis," *Acta Biomater.*, vol. 98, pp. 36–49, 2019, doi: 10.1016/j.actbio.2019.05.056.

[24]  Ahmed, M., Lorimer, G. W., Lyon, P. and Pilkington, R., "The effect of heat treatment and composition on the microstructure and properties of cast Mg-Y-RE alloys," in *Proc. Magnesium Alloys and Their Applications*, ed. B. L. Mordike and F. Hehmann. DGM Informationsgesellschaft, Germany, 1992, p. 301.

[25]  M. P. Staiger, A. M. Pietak, J. Huadmai, G. Dias, Magnesium and its alloys as orthopedic biomaterials: a review, Biomaterials 27 (2006) 1728–1734.

[26]  J. F. Nie and B. C. Muddle, "Characterisation of strengthening precipitate phases in a Mg-Y-Nd alloy," *Acta Mater.*, vol. 48, no. 8, pp. 1691–1703, 2000, doi: 10.1016/S1359-6454(00)00013-6.

[27]  P. J. Apps, H. Karimzadeh, J. F. King, and G. W. Lorimer, "Precipitation reactions in magnesium-rare earth alloys containing yttrium, gadolinium or dysprosium," *Scr. Mater.*, vol. 48, no. 8, pp. 1023–1028, 2003, doi: 10.1016/S1359-6462(02)00596-1.

[28]  C. Antion, P. Donnadieu, F. Perrard, A. Deschamps, C. Tassin, and A. Pisch, "Hardening precipitation in a Mg-4Y-3RE alloy," *Acta Mater.*, vol. 51, no. 18, pp. 5335–5348, 2003, doi: 10.1016/S1359-6454(03)00391-4.

[29]  H. Windhagen, K. Radtke, A. Weizbauer, J. Diekmann, *et al*. "Biodegradable magnesium-based screw clinically equivalent to titanium screw in hallux valgus surgery: short term results of the first prospective, randomized, controlled clinical pilot study." *BioMed Eng OnLine* 12, 62 (2013).

[30]  M. Haude, H. Ince, A. Abizaid, *et al*. "Sustained safety and performance of the second-generation drug-eluting absorbable metal scaffold in patients with de novo coronary lesions: 12-month clinical results and angiographic findings of the BIOSOLVE-II first-in-man trial," European Heart Journal, vol. 37, Issue 35, pp. 2701–2709, 2016.

[31]  M. Marvi-Mashhadi, C. S. Lopes, J. LLorca. High fidelity simulation of the mechanical behavior of closed-cell polyurethane foams. Journal of the Mechanics and Physics of Solids, 135, 103814, 2020.

[32]  J.-F. Nie, Physical Metallurgy of Light Alloys. In: Laughlin, D.E., Hono, K. (Eds.), Physical Metallurgy, pp. 2009–2156 (2014).

[33]   V. Herrera-Solaz, J. LLorca, E. Dogan, I. Karaman, J. Segurado. An inverse optimization strategy to determine single crystal mechanical behavior from polycrystal tests: application to AZ31 Mg alloy. International Journal of Plasticity, 57, 1-15, 2014.

[34]   P. Hidalgo-Manrique, V. Herrera-Solaz, J. Segurado, J. LLorca, F. Gálvez, O. A. Ruano, S. Yi, M. T. Pérez-Prado. Origin of the reversed yield asymmetry in Mg-rare earth alloys at high temperature. Acta Materialia, 92, 265–277, 2015.

[35]   C. M. Cepeda-Jiménez, J. M. Molina-Aldareguia, M. T. Pérez-Prado. Effect of grain size on slip activity in pure magnesium polycrystals, Acta Materialia 84, 443-456, 2015.

[36]   D. Gianoglio *et al.*, "Banded microstructures in rapidly solidified Al-3 wt% Er," *Intermetallics*, 119, 106724, 2020, doi: 10.1016/j.intermet.2020.106724.

[37]   J. T. McKeown *et al.*, "In situ transmission electron microscopy of crystal growth-mode transitions during rapid solidification of a hypoeutectic Al–Cu alloy," *Acta Mater.*, vol. 65, pp. 56–68, 2014, doi: https://doi.org/10.1016/j.actamat.2013.11.046.

[38]   M. Zimmermann, M. Carrard, M. Gremaud, and W. Kurz, "Characterization of the banded structure in rapidly solidified Al Cu alloys," *Mater. Sci. Eng. A*, vol. 134, pp. 1278–1282, 1991, doi: https://doi.org/10.1016/0921-5093(91)90973-Q.

[39]   M. Carrard, M. Gremaud, M. Zimmermann, and W. Kurz, "About the banded structure in rapidly solidified dendritic and eutectic alloys," *Acta Metall. Mater.*, vol. 40, no. 5, pp. 983–996, 1992, doi: https://doi.org/10.1016/0956-7151(92)90076-Q.

[40]   W. J. Boettinger, D. Shechtman, R. J. Schaefer, and F. S. Biancaniello, "The Effect of Rapid Solidification Velocity on the Microstructure of Ag-Cu Alloys," *Metall. Trans. A*, vol. 15, no. 1, pp. 55–66, 1984, doi: 10.1007/BF02644387.

[41]   W. Kurz and R. Trivedi, "Overview No. 87 Solidification microstructures: Recent developments and future directions," *Acta Metall. Mater.*, vol. 38, no. 1, pp. 1–17, 1990, doi: https://doi.org/10.1016/0956-7151(90)90129-5.

[42]   M. Gremaud, M. Carrard, and W. Kurz, "Banding phenomena in Al-Fe alloys subjected to laser surface treatment," *Acta Metall. Mater.*, vol. 39, no. 7, pp. 1431–1443, 1991, doi: https://doi.org/10.1016/0956-7151(91)90228-S.

[43]   S. R. Coriell and R. F. Sekerka, "Oscillatory morphological instabilities due to non-equilibrium segregation," *J. Cryst. Growth*, vol. 61, no. 3, pp. 499–508, 1983, doi: https://doi.org/10.1016/0022-0248(83)90179-3.

[44]   M. J. Aziz and T. Kaplan, "Continuous growth model for interface motion during alloy solidification," *Acta Metall.*, vol. 36, no. 8, pp. 2335–2347, 1988, doi: https://doi.org/10.1016/0001-6160(88)90333-1.

[45]   S. Thomas, N. V Medhekar, G. S. Frankel, and N. Birbilis, "Corrosion mechanism and hydrogen evolution on Mg," *Curr. Opin. Solid State Mater. Sci.*, vol. 19, no. 2, pp. 85–94, 2015, doi: https://doi.org/10.1016/j.cossms.2014.09.005.

[46]   Z. Li, X. Gu, S. Lou, and Y. Zheng, "The development of binary Mg–Ca alloys for use as biodegradable materials within bone," Biomaterials, vol. 29, no. 10, pp. 1329–1344, 2008, doi: https://doi.org/10.1016/j.biomaterials.2007.12.021.

[47]   W. Ding, "Opportunities and challenges for the biodegradable magnesium alloys as next-generation biomaterials," *Regen. Biomater.*, vol. 3, no. 2, pp. 79–86, Mar. 2016, doi:

10.1093/rb/rbw003.

[48] C. Shuai *et al.*, "Nd-induced honeycomb structure of intermetallic phase enhances the corrosion resistance of Mg alloys for bone implants," *J. Mater. Sci. Mater. Med.*, vol. 28, no. 9, p. 130, 2017, doi: 10.1007/s10856-017-5945-0.

[49] N. T. Kirkland, J. Lespagnol, N. Birbilis, and M. P. Staiger, "A survey of bio-corrosion rates of magnesium alloys," *Corros. Sci.*, vol. 52, no. 2, pp. 287–291, 2010, doi: https://doi.org/10.1016/j.corsci.2009.09.033.

[50] J. Parthasarathy, B. Starly, S. Raman, and A. Christensen, "Mechanical evaluation of porous titanium (Ti6Al4V) structures with electron beam melting (EBM)," *J. Mech. Behav. Biomed. Mater.*, vol. 3, no. 3, pp. 249–259, 2010, doi: https://doi.org/10.1016/j.jmbbm.2009.10.006.

[51] J.-Y. Rho, L. Kuhn-Spearing, and P. Zioupos, "Mechanical properties and the hierarchical structure of bone," *Med. Eng. Phys.*, vol. 20, no. 2, pp. 92–102, 1998, doi: https://doi.org/10.1016/S1350-4533(98)00007-1.

[52] K. A. Athanasiou, C.-F. Zhu, D. R. Lanctot, C. M. Agrawal, and X. Wang, "Fundamentals of Biomechanics in Tissue Engineering of Bone," *Tissue Eng.*, vol. 6, no. 4, pp. 361–381, Aug. 2000, doi: 10.1089/107632700418083.

[53] H. Li, J. Wen, Y. Liu, J. He, H. Shi, P. Tian, " Progress in Research on Biodegradable Magnesium Alloys: A Review," *Adv. Engng. Mater.*, vol. 22, no. 7, 2000213, 2020, doi: 10.1002/adem.202000213.

[54] K. Lietaert, A. A. Zadpoor, M. Sonnaert, J. Schrooten, L.Weber, A. Mortensen, J.Vleugels, " Mechanical properties and cytocompatibility of dense and porous Zn produced by laser powder b e d fusion for biodegradable implant applications," *Acta Biomateriala,* vol. 110, pp. 289-302, 2020, doi: 10.1016/j.actbio.2020.04.006

[55] S. Agarwal, J. Curtin, B. Duff, S. Jaiswal, "Biodegradable magnesium alloys for orthopaedic applications: A review on corrosion, biocompatibility and surface modifications," *Materials Science and Engineering: C*, vol. 68, pp. 948-963, 2020.

## SUPPLEMENTARY MATERIAL

**Video S1:** Successive cross-sections of the tomogram of the WE43 Mg alloy scaffold with nominal strut diameter of 750 μm.

**Video S2:** Successive cross-sections of the tomogram of the WE43 Mg alloy scaffold with nominal strut diameter of 500 μm.

**Video S3:** Successive cross-sections of the tomogram of the WE43 Mg alloy scaffold with nominal strut diameter of 250 μm.

**Video S4:** Optical video of the compressive deformation of the WE43 Mg alloy scaffold with nominal strut diameter of 750 μm.

**Video S5:** Optical video of the compressive deformation of the WE43 Mg alloy scaffold with nominal strut diameter of 250 μm.

**Video S6:** Optical video of the compressive deformation of the WE43 Mg alloy scaffold with nominal strut diameter of 500 μm in the T4 condition.

**Video S7:** Successive cross-sections of the tomogram of the WE43 Mg alloy scaffold with nominal strut diameter of 500 μm in the as-printed condition after 7 days of immersion in DMEM.

**Video S8:** Successive cross-sections of the tomogram of the WE43 Mg alloy scaffold with nominal strut diameter of 500 μm in the T4 condition after 7 days of immersion in DMEM.